\documentclass[11pt]{article}
\usepackage{latexsym}
\usepackage{times}
\usepackage{graphicx}

\newcommand{\one}{\mbox{\bf 1}}
\newcommand{\zero}{\mbox{\bf 0}}

\newcommand{\fname}[1]{\mbox{\it #1}}

\newcommand{\lo}{\fname{lo}}
\newcommand{\hi}{\fname{hi}}

\newcommand{\opname}[1]{\mbox{\sc #1}}

\newcommand{\andop}{\opname{and}}
\newcommand{\orop}{\opname{or}}
\newcommand{\xorop}{\opname{xor}}
\newcommand{\iteop}{\opname{ITE}}

\newcommand{\applyop}{\opname{apply}}

\newcommand{\dctype}{\textsc{don't-care}}
\newcommand{\ortype}{\textsc{or}}
\newcommand{\ztype}{\textsc{zero}}

\newcommand{\lpair}[2]{#1 \mathbin{\rm :} #2}
\newcommand{\bnode}[3]{\left \langle #1 \rightarrow #3 , #2 \right \rangle}
\newcommand{\znode}[3]{\left \langle #1 \rightarrow #3 , #2 \right \rangle}

\newcommand{\msizeratio}[3]{R_{#1}(#2, #3)}
\newcommand{\sizeratio}[3]{\msizeratio{#1}{\rm #2}{\rm #3}}

\begin{document}

\title{Chain Reduction for \\Binary and Zero-Suppressed Decision Diagrams}

\author{Randal E. Bryant \\
Computer Science Department \\
Carnegie Mellon University\\
{\tt Randy.Bryant@cs.cmu.edu}}

\maketitle

\begin{abstract}
Chain reduction enables reduced ordered binary decision diagrams
(BDDs) and zero-suppressed binary decision diagrams (ZDDs) to each
take advantage of the others' ability to symbolically represent
Boolean functions in compact form.  For any Boolean function, its
chain-reduced ZDD (CZDD) representation  will be no larger than
its ZDD representation, and at most twice the size of its BDD
representation.  The chain-reduced BDD (CBDD) of a function will be no larger
than its BDD representation, and at most three times the size of its
CZDD representation.  Extensions to the standard algorithms for
operating on BDDs and ZDDs enable them to operate on the chain-reduced
versions.  Experimental evaluations on representative benchmarks for
encoding word lists, solving combinatorial problems, and operating
on digital circuits indicate that chain reduction can provide
significant benefits in terms of both memory and execution time.
\end{abstract}

\section{Introduction}

Decision diagrams (DDs) encode sets of values
in compact forms, such that operations on the sets can be performed on
the encoded representation, without expanding the sets into their
individual elements.  In this paper, we consider two classes of
decision diagrams: reduced ordered binary decision diagrams (BDDs)
\cite{bryant:tc86} and zero-suppressed binary decision diagrams (ZDDs)
\cite{minato:dac93,minato:book95}.  These two representations are
closely related to each other, with each achieving more compact
representations for different classes of applications.  We present
extensions to both representations, such that BDDs can take
advantage of the source of compaction provided by ZDDs, and
vice-versa.

Both BDDs and ZDDs encode sets of binary sequences of some fixed
length $n$, defining a Boolean function over $n$ variables.  We can
bound their relative sizes as follows.  Suppose for some function, we
encode it according to the different DD types.  For
function $f$, let $T(f)$ indicate the number of nodes (including leaf
nodes) in the representation of type $T$.  Let
$\msizeratio{f}{T_1}{T_2}$ denote the relative sizes when representing
$f$ using types $T_1$ and $T_2$:
\begin{displaymath}
\begin{array}{lcl}
\msizeratio{f}{T_1}{T_2} & = & \frac{T_1(f)}{T_2(f)}
\end{array}
\end{displaymath}

Comparing BDDs and ZDDs, Knuth \cite{knuth:v4a} has shown that for any
function $f$:
\begin{eqnarray}
\sizeratio{f}{BDD}{ZDD} & \leq & n/2 + o(n) \label{eqn:bdd-zdd} \\
\sizeratio{f}{ZDD}{BDD} & \leq & n/2 + o(n) \label{eqn:zdd-bdd}
\end{eqnarray}
These bound improve on the size ratios of $n$ derived by Wegener \cite{wegener:book00}.

As these bounds show, ZDDs may be significantly (a factor of $n/2$)
more compact than BDDs, or vice-versa.  Hence, it can be critical to
choose one representation over another in a given application.  Indeed,
users of DDs often switch between the two representations during a
sequence of operations in order take advantage of their different characteristics.

\begin{figure}
\begin{center}
\includegraphics[scale=0.60]{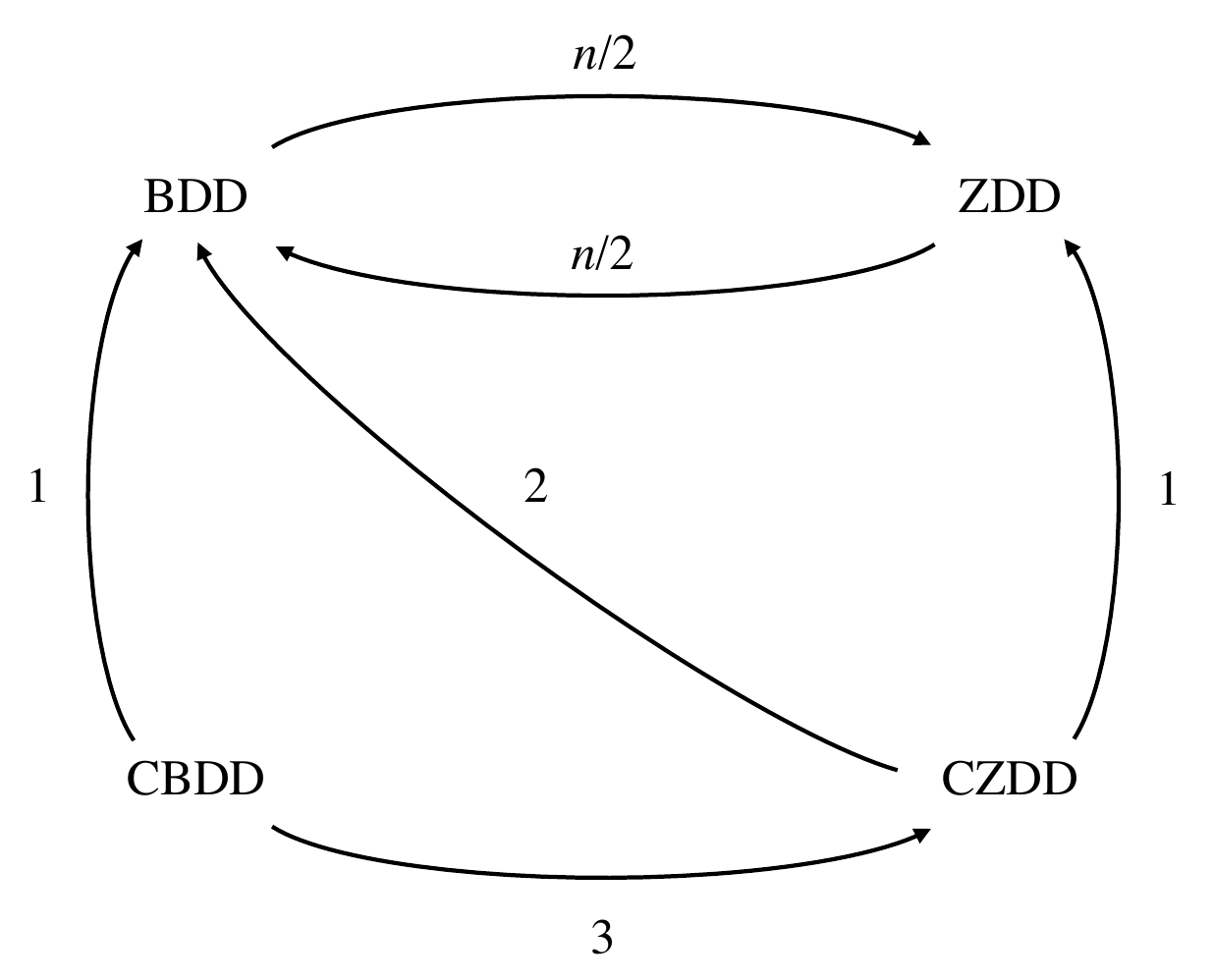}
\end{center}
\caption{{\bf Size bound relations between different representations}}
\label{fig:relations}
\end{figure}

In this paper, we introduce two new representations: {\em
  chain-reduced ordered binary decision diagrams} (CBDDs), and {\em
  chain-reduced zero-suppressed binary decision diagrams} (CZDDs).
The key idea is to associate two levels with each node and to use such
nodes to encode particular classes of linear chains found in BDDs
and ZDDs.
Chain reduction can be defined in terms of a set of
reduction rules applied to BDDs and ZDDs,
giving bounds for any function $f$
\begin{eqnarray}
\sizeratio{f}{CBDD}{BDD} & \leq & 1 \label{eqn:cbdd-bdd} \\
\sizeratio{f}{CZDD}{ZDD} & \leq & 1   \label{eqn:czdd-zdd} 
\end{eqnarray}
In terms of graph sizes, using chain reduction can only lead to
more compact representations.

We show bounds on the relative sizes of the representations as:
\begin{eqnarray}
\sizeratio{f}{CBDD}{CZDD} & \leq & 3 + o(1) \label{eqn:cbdd-czdd} \\
\sizeratio{f}{CZDD}{BDD} & \leq & 2 + o(1) \label{eqn:czdd-bdd} 
\end{eqnarray}
These relations are summarized in the diagram of Figure
\ref{fig:relations}.  In this figure, each arc from type $T_1$ to type
$T_2$ labeled by an expression $E$ indicates that
$\msizeratio{f}{T_1}{T_2} \leq E + o(E)$. These arcs define a transitive
relation, and so we can also infer that
\begin{eqnarray}
\sizeratio{f}{CBDD}{ZDD} & \leq & 3 + o(1) \label{eqn:cbdd-zdd} 
\end{eqnarray}
These results indicate that the two compressed representations
will always be within a small constant factor (2 for CZDDs and 3 for
CBDDs) of either a BDD or a ZDD representation.
While one representation may be more slightly compact than the other,
the relative advantage is bounded by a constant factor, and hence
choosing between them is less critical.

This paper defines the two compressed representations, derives the
bounds indicated in Equations \ref{eqn:cbdd-czdd}--\ref{eqn:czdd-bdd}
and presents extensions of the core BDD and ZDD algorithms to their
chained versions.  It describes an implementation based on
modifications of the CUDD BDD package \cite{somenzi:sttt01}.  It
presents some experimental results and concludes with a discussion of
the merits of chaining and possible extensions.

\section{BDDs and ZDDs}

Both BDDs and ZDDs encode sets of binary sequences of length $n$
as directed acyclic graphs with two leaf nodes, labeled
with values $\zero$ and $\one$, which we refer to as ``leaf $\zero$'' and ``leaf $\one$,'' respectively.
Each nonleaf node $v$ has an
associated level $l$, such that $1 \leq l \leq n$, and two outgoing
edges, labeled $\lo$ and $\hi$ to either a leaf node or a nonleaf
node. 
By convention, leaf nodes have level $n+1$. 
An edge from $v$ to node $u$ having level $l'$ must have $l < l'$.  

\begin{figure}[tb]
\begin{center}
\begin{picture}(6,3.54)
\put(0,3.34){(A) Levelized BDD}
\put(0.04,0){\includegraphics[scale=0.70]{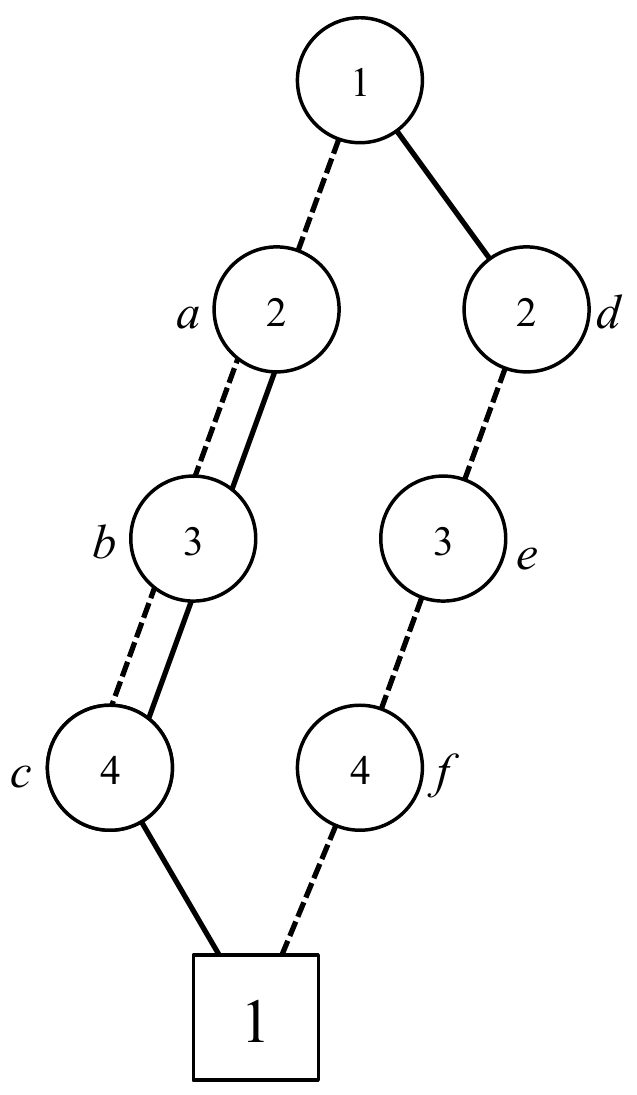}}
\put(2.00,3.34){(B) BDD}
\put(2.09,0){\includegraphics[scale=0.70]{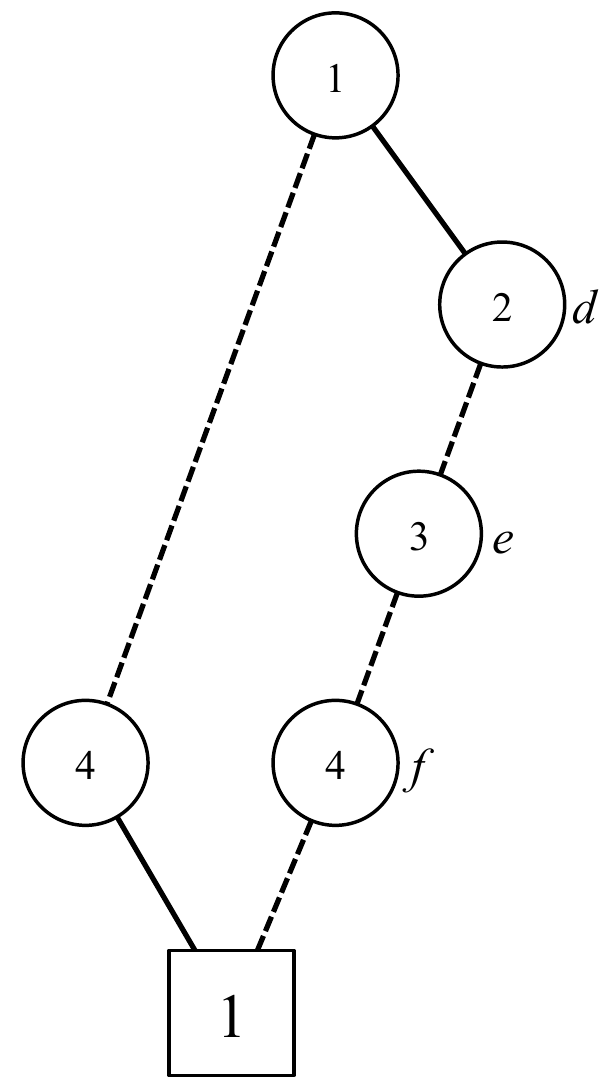}}
\put(4.0,3.34){(C) ZDD}
\put(4.34,0){\includegraphics[scale=0.70]{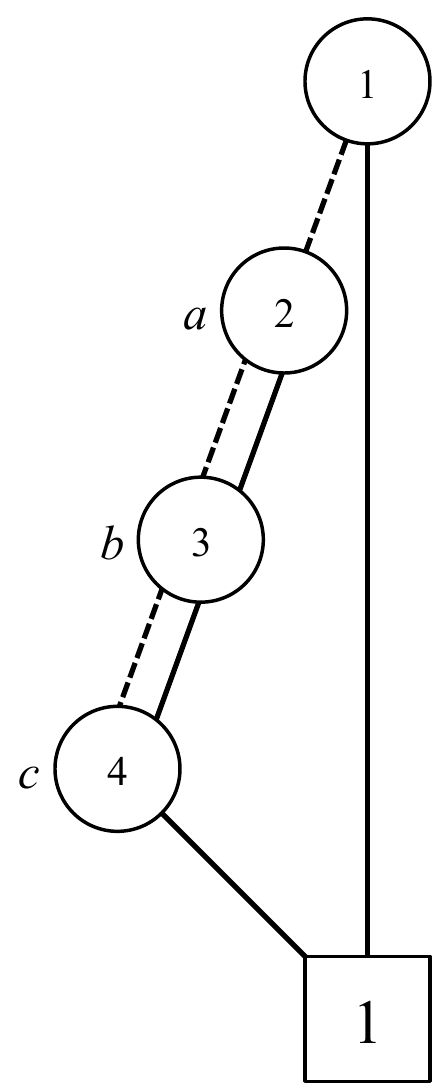}}
\end{picture}
\end{center}
\caption{{\bf Reductions in BDDs and ZDDs}.  Each reduces the representation size with with edges between nonconsecutive levels.}
\label{fig:bdd-zdd-eg}
\end{figure}

Figure \ref{fig:bdd-zdd-eg} shows three decision-diagram
representations of the set $S$, defined as:
\begin{eqnarray}
S & = & \{ 0001, 0011, 0101, 0111, 1000\}
\label{eqn:set-example}
\end{eqnarray}
The $\lo$ edge from each node is
shown as a dashed line, and the $\hi$ edge is shown as a solid line. 
As a shorthand, we omit leaf $\zero$ and all branches to it.  

Graph A represents $S$ as a {\em levelized binary decision diagram},
where an edge from a node with level $l$ must connect to either
leaf $\zero$ or to a node with level $l+1$.  (This is
similar to the {\em quasi-reduced} form described by Knuth
\cite{knuth:v4a}, except that he only allows edges to leaf
$\zero$ from nodes at level $n$.)  Each path from the root to
leaf $\one$ encodes an element of set $S$.  For a
given path, the represented sequence has value 0 at position $l$ when
the path follows the $\lo$ edge from the node with level $l$ and
value 1 when the path follows the $\hi$ edge.

Graph A has nodes forming two linear chains: a {\em \dctype{} chain},
consisting of nodes $a$ and $b$, and an {\em \ortype{} chain},
consisting of nodes $d$, $e$, and $f$.  A \dctype{} chain is
a series of \dctype{} nodes, each having
its two outgoing edges directed to the same next node.
In terms of the set of represented binary sequences, a \dctype{} node with level $l$
allows both values 0 and 1 at sequence position $l$.  An \ortype{} chain
consists of a sequence where the outgoing $\hi$ edges for the
nodes all go the same node---in this case, leaf $\zero$.  An \ortype{} chain
where all $\hi$ edges lead to value $\zero$ has only a single
path, assigning value 0 to the corresponding positions in the
represented sequence.  We will refer to this special class of \ortype{} chain as a {\em \ztype{} chain}.

BDDs and ZDDs differ from each other in the interpretations they
assign to a {\em level-skipping edge}, when a node with level $l$ has
an edge to a node with level $l'$ such that $l + 1 < l'$.  For BDDs,
such an edge is considered to encode a \dctype{} chain.  Thus, graph B
in Figure \ref{fig:bdd-zdd-eg} shows an BDD encoding set $S$.  The
edge on the left from level 1 to level 4 is equivalent to the
\dctype{} chain formed by nodes $a$ and $b$ of graph A\@.  For ZDDs, a
level skipping edge encodes a \ztype{} chain.  Thus, graph C shows a
ZDD encoding set $S$.  The edge on the right from level 1 to the leaf
encodes the \ztype{} chain formed by nodes $d$, $e$, and $f$ of graph
A\@.  Whether the set is encoded as a BDD or a ZDD, one type of
linear chains remains.  Introducing chain reduction enables BDDs
and ZDDs to exploit both \dctype{} and \ortype{} (and
therefore \ztype{}) chains to compress their representations.

\section{Chain Patterns and Reductions}

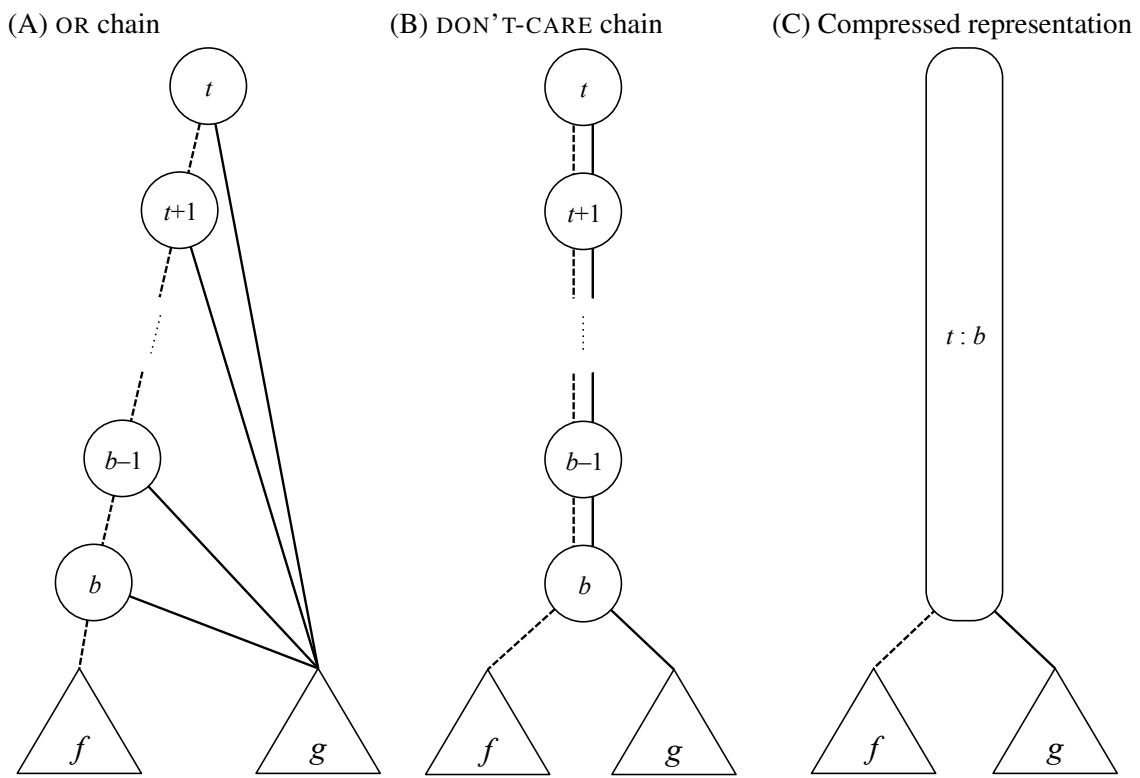
\begin{figure}
\begin{center}
\begin{picture}(6.00,4.10)
\put(0,3.90){(A) \ortype{} chain}
\put(0.02,0){\includegraphics[scale=0.60]{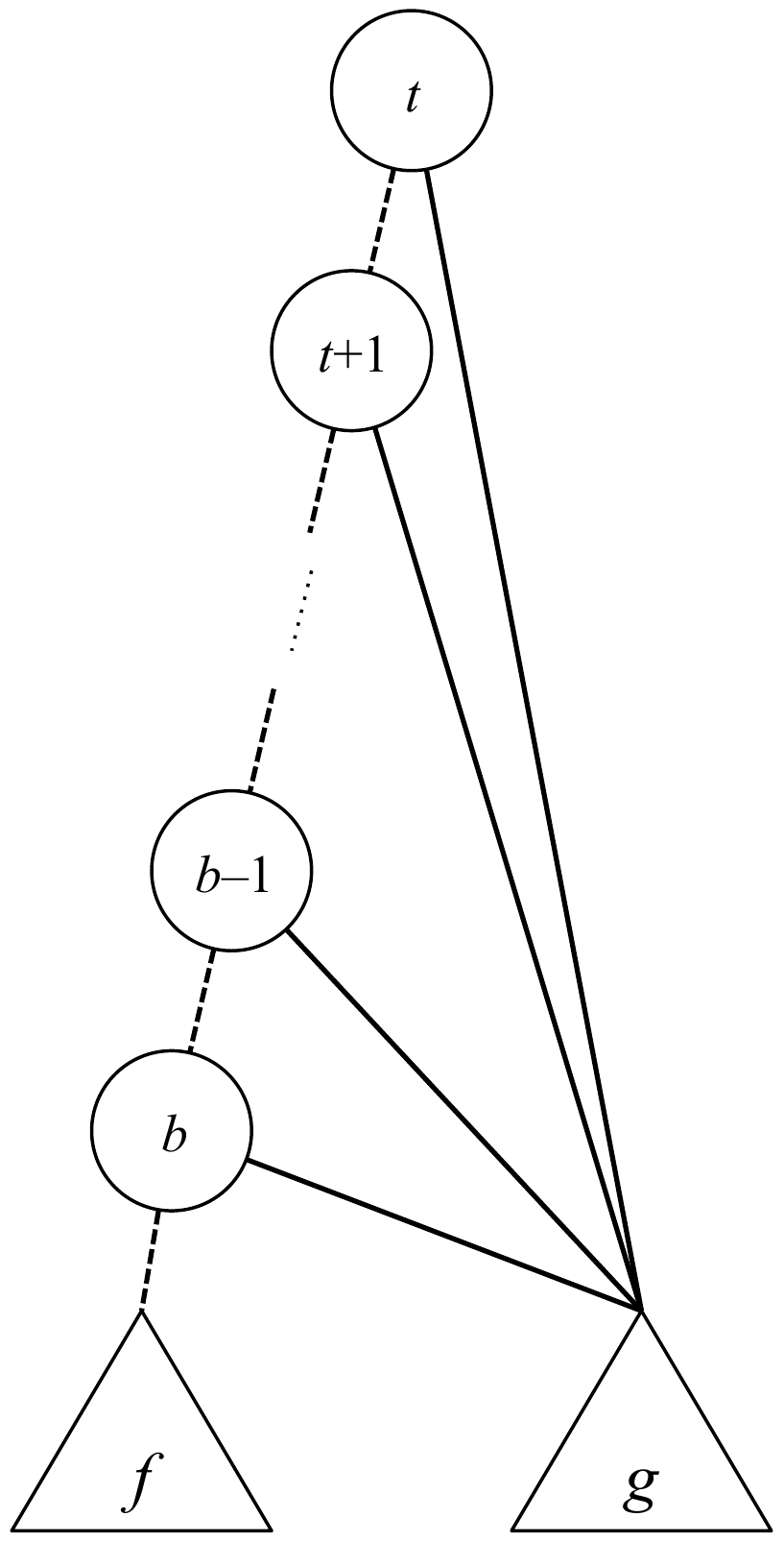}}
\put(2.0,3.90){(B) \dctype{} chain}
\put(2.16,0){\includegraphics[scale=0.60]{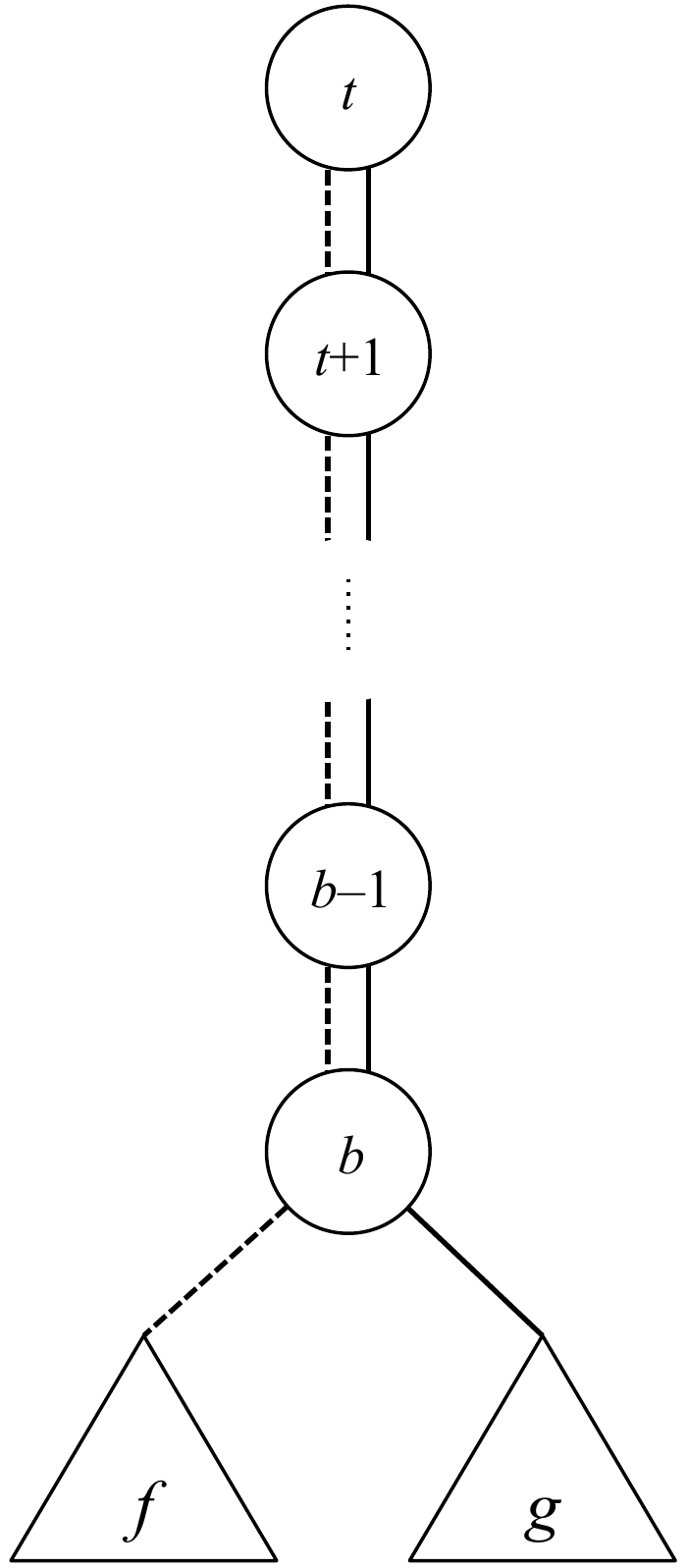}}
\put(4.0,3.90){(C) Compressed representation}
\put(4.18,0){\includegraphics[scale=0.60]{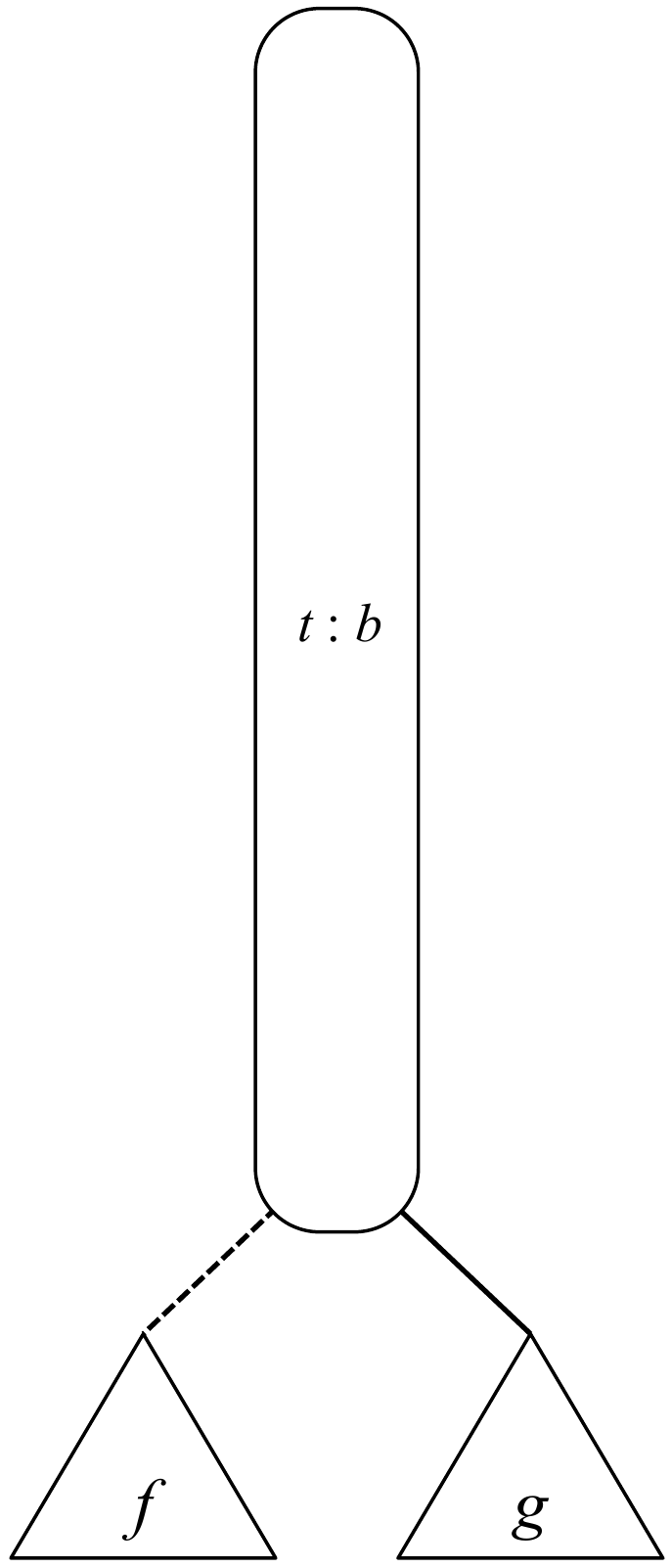}}
\end{picture}
\end{center}
\caption{{\bf Chain patterns}.  These patterns remain after BDD reduction (A), and ZDD reduction (B), but can be represented in compressed form (C).}
\label{fig:chain-patterns}
\end{figure}

Figure \ref{fig:chain-patterns} shows the general form of \ortype{} and \dctype{} chains, as were illustrated in the examples of Figure
\ref{fig:bdd-zdd-eg}.  These chains have
 levels ranging from $t$ to $b$, such that
$1 \leq t < b \leq n$.  Each form consists of a linear chain of nodes
followed by two nodes $f$ and $g$ with levels greater than $b$.
Nodes $f$ and $g$ are drawn as triangles to indicate that they are the
roots of two subgraphs in the representation.
In an
\ortype{} chain, the $\lo$ edge from each node is directed to the next node
in the chain, and the $\hi$ edge is directed to node $g$.  The chains
eliminated by ZDDs are a special case where $g = \zero$.
In a
\dctype{} chain, both the $\lo$ and the $\hi$ edges are directed
toward the next node in the chain.

As was illustrated in Figure \ref{fig:bdd-zdd-eg}, having edges that
skip levels allows BDDs to compactly represent \dctype{} chains and ZDDs to eliminate
\ortype{} chains when $g = \zero$.  The goal of chain
reduction is to allow both forms to compactly represent both types of chains.
They do so by associating two levels with each node, as indicated in
Figure \ref{fig:chain-patterns}(C).  That is, every nonleaf node has an associated pair
of levels $\lpair{t}{b}$, such that $1 \leq t \leq b \leq n$.  In a
{\em chain-reduced ordered binary decision diagram} (CBDD), such a
node encodes the \ortype{} chain pattern shown in Figure
\ref{fig:chain-patterns}(A), while in a {\em chain-reduced
  zero-suppressed binary decision diagram} (CZDD), such a node encodes
the \dctype{} chain pattern shown in Figure
\ref{fig:chain-patterns}(B).  A node with levels $t$ and $b$ such
that $t=b$ encodes a standard node with respect to the indicated variable.

\begin{figure}
\begin{center}
\begin{picture}(6.3,3.54)
\put(0,3.34){(A) Levelized BDD}
\put(0.09,0){\includegraphics[scale=0.75]{eg-a}}
\put(2.00,3.34){(B) CBDD}
\put(2.12,0){\includegraphics[scale=0.75]{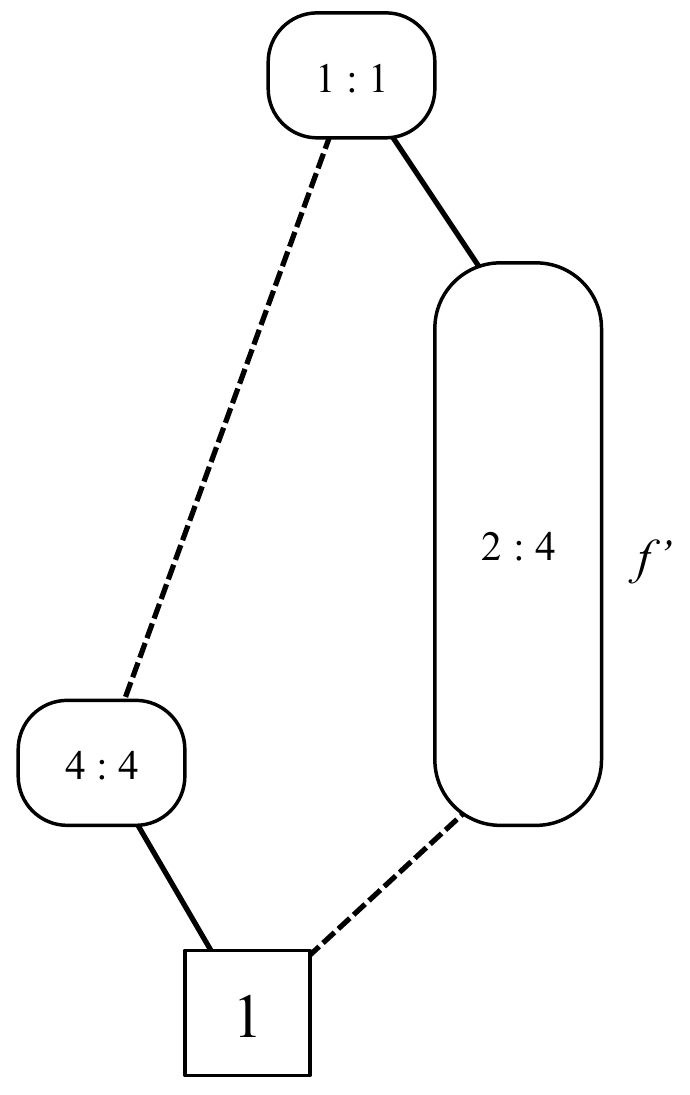}}
\put(4.0,3.34){(C) CZDD}
\put(4.64,0){\includegraphics[scale=0.75]{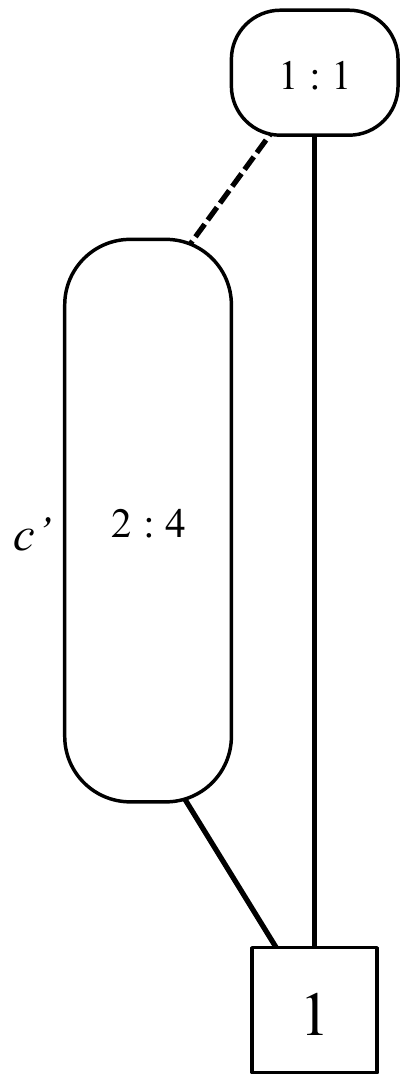}}
\end{picture}
\end{center}
\caption{{\bf Chain Reduction Examples}.  Each now reduces both chain types.}
\label{fig:bdd-zdd-reduce-eg}
\end{figure}

Figure \ref{fig:bdd-zdd-reduce-eg} shows the effect of chain
reduction, starting with the levelized graph A\@.
In the CBDD (B), a single node $f'$ replaces the \ortype{} chain consisting
of nodes $d$, $e$, and $f$.  In the CZDD (C), the \dctype{} chain
consisting of nodes $a$ and $b$ is incorporated into node $c$ to form
node $c'$.  These new nodes are drawn in elongated form to emphasize that
they span a range of levels, but it should be emphasized that
{\em all} nodes in a chained representation have an associated pair of levels.

\section{Generating CBDDs and CZDDs}

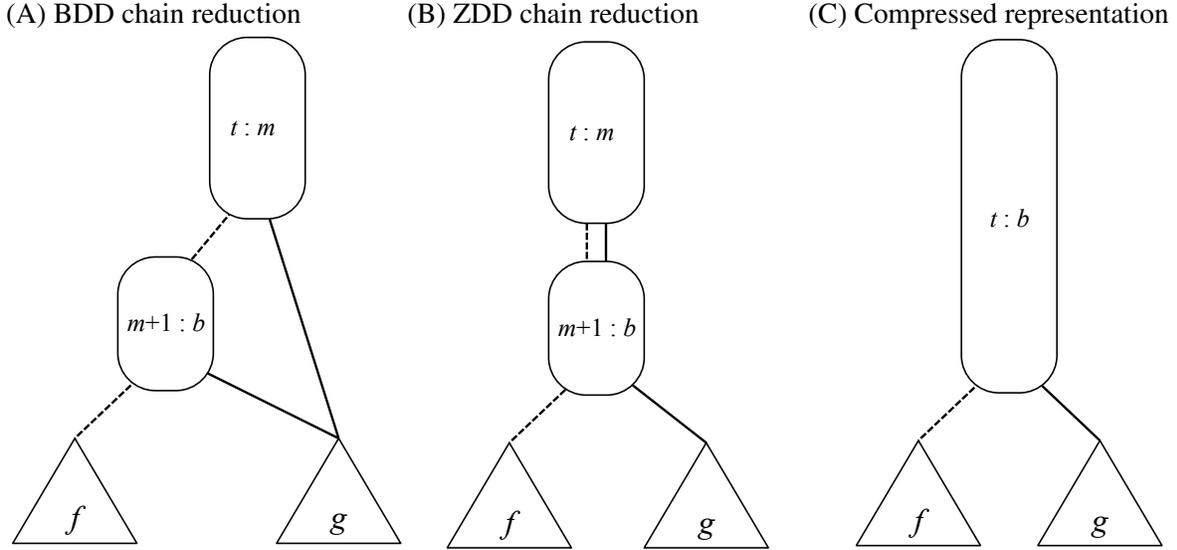
\begin{figure}
\begin{center}
\begin{picture}(6.30,2.95)
\put(0,2.75){(A) BDD chain reduction}
\put(0.01,0){\includegraphics[scale=0.60]{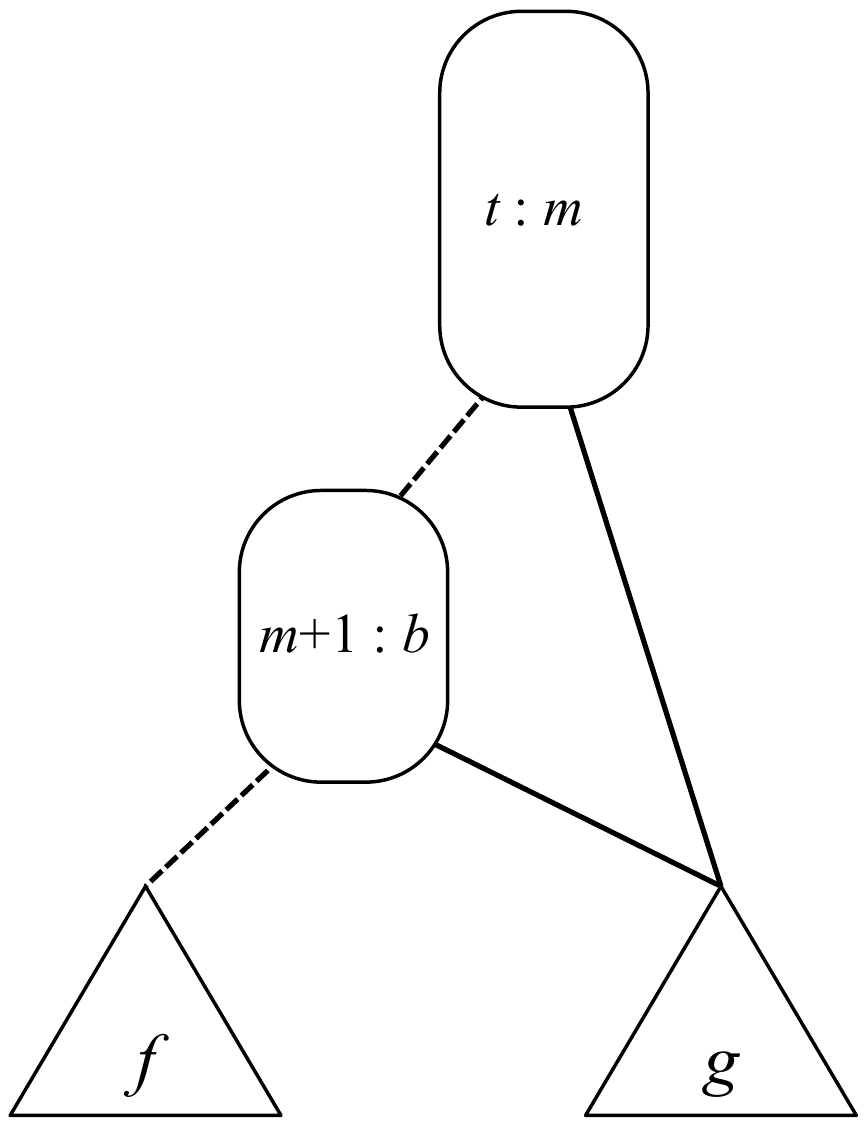}}
\put(2.10,2.75){(B) ZDD chain reduction}
\put(2.29,0){\includegraphics[scale=0.60]{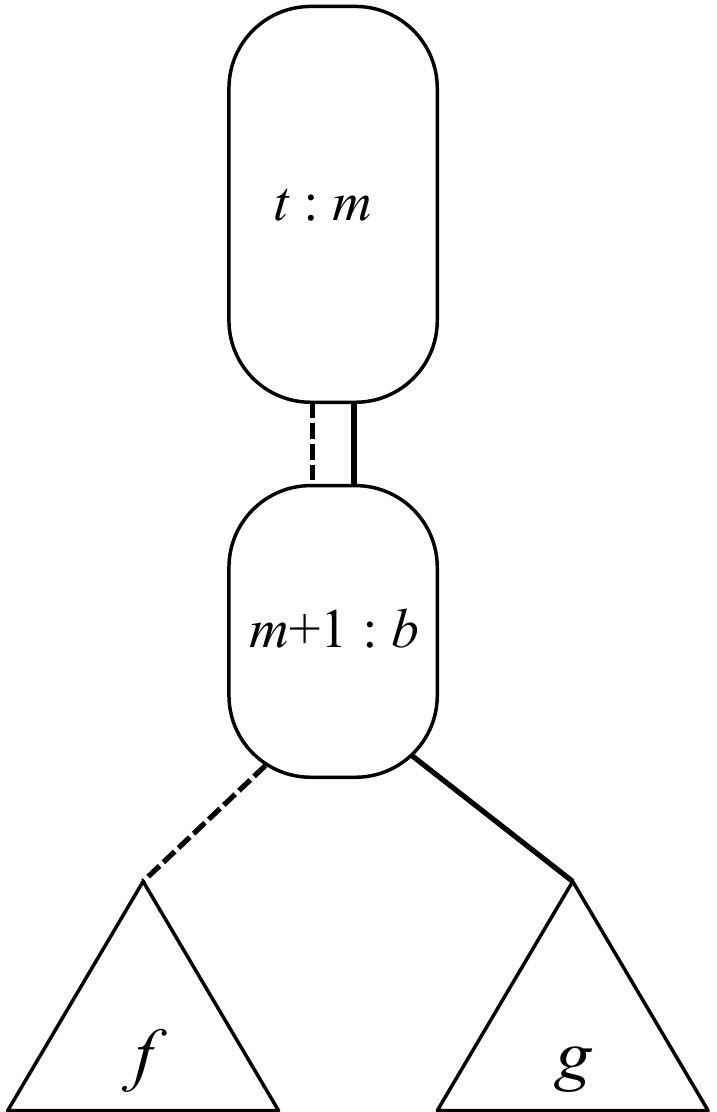}}
\put(4.20,2.75){(C) Compressed representation}
\put(4.43,0){\includegraphics[scale=0.60]{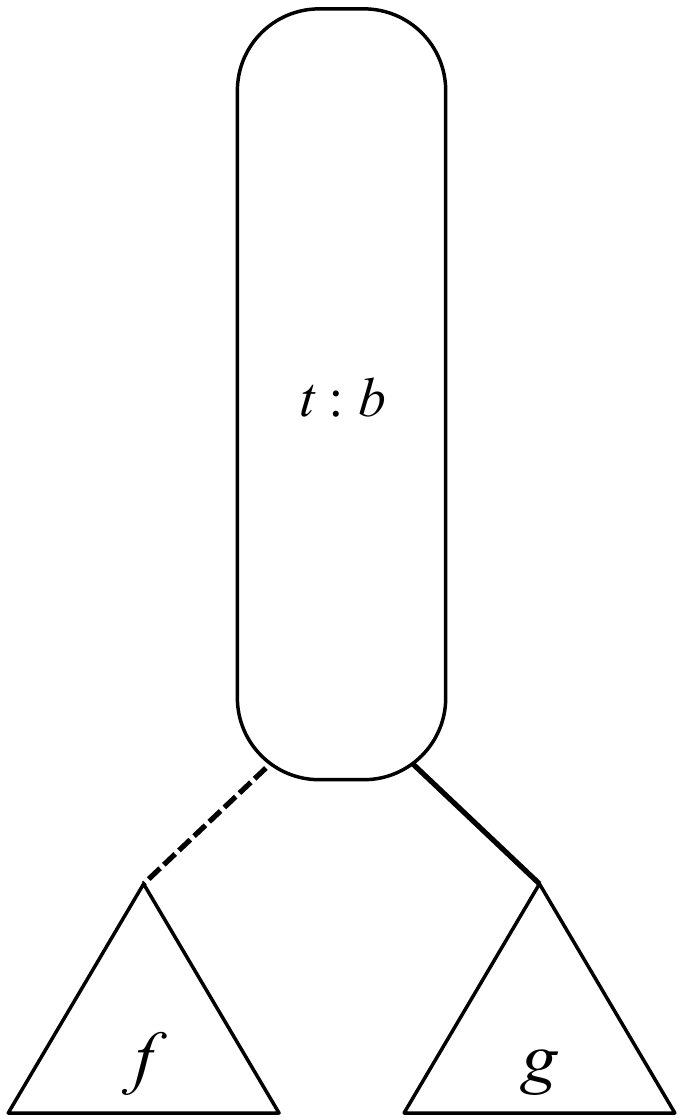}}
\end{picture}
\end{center}
\caption{{\bf Chain reduction cases}.  These cases define how chain reduction can be applied to BDDs (A) and ZDDs to obtain the single chain node (C).}
\label{fig:chain-reduction}
\end{figure}

As notation, let us denote a node of the form illustrated in Figure
\ref{fig:chain-patterns}(C) with the modified if-then-else notation
$\bnode{\lpair{t}{b}}{f}{g}$.  That is, the node has a range of levels
from $t$ to $b$, an outgoing $\hi$ edge to node $g$, and an outgoing
$\lo$ edge to node $f$.

A BDD representation of a function can be transformed into a CBDD as follows.
The process starts by labeling each
node having level $l$ in the BDD with the pair $\lpair{t}{b}$, such that $t = b = l$.
Then, we repeatedly apply a {\em reduction rule}, replacing any pair of nodes $u$ and $v$ of the form
$u = \bnode{\lpair{t}{m}}{v}{g}$ and $v =
\bnode{\lpair{m+1}{b}}{f}{g}$ (illustrated in Figure
\ref{fig:chain-reduction}(A))
by the single node
$\bnode{\lpair{t}{b}}{f}{g}$ (illustrated in Figure
\ref{fig:chain-reduction}(C)).
A similar process can transfrom any ZDD representation of a function 
into a CZDD, using the reduction rule that a pair of nodes $u$ and $v$ of the form
$u = \bnode{\lpair{t}{m}}{v}{v}$ and $v =
\bnode{\lpair{m+1}{b}}{f}{g}$ 
(illustrated in Figure
\ref{fig:chain-reduction}(B))
is replaced by the single node
$\bnode{\lpair{t}{b}}{f}{g}$ (illustrated in Figure
\ref{fig:chain-reduction}(C)).
In practice, most algorithms for constructing decision diagrams
operate from the bottom up.  The reduction rules are applied as nodes are created, and so
unreduced nodes are never actually generated.

\section{Size Ratio Bounds}

These reduction rules allows us to bound the relative sizes of the
different representations, as given by Equations \ref{eqn:cbdd-czdd} and \ref{eqn:czdd-bdd}.

First, let us consider Equation
\ref{eqn:cbdd-czdd}, bounding the relative sizes of the CBDD and CZDD
representations of a function.  Consider a graph $G$ representing
function $f$ as a CZDD\@.
We can generate a CBDD representation $G'$ as follows.  $G'$ contains
a node $v'$ for each node $v$ in $G$.  However, if $v$ has levels
$\lpair{t}{b}$, then $v'$ has levels $\lpair{b}{b}$, because any
\dctype{} chain encoded explicitly in the CZDD is encoded implicitly
in a CBDD\@.  

Consider an edge from node $u$ to node $v$ in $G$, where the nodes
have levels $\lpair{t_u}{b_u}$ and $\lpair{t_v}{b_v}$, respectively.  If $t_v
= b_u+1$, then there can be an edge directly from $u'$ to $v'$.  If $t_v
< b_u+1$, then we introduce a new node to encode the implicit zero
chain in $G$ from $u$ to $v$.  This node
has the form
$\bnode{\lpair{b_u+1}{t_v-1}}{v'}{\zero}$ and has an edge from $u'$ to it.

The size of $G'$ is bounded by the number of nodes plus the number of
edges in $G$.  Since each node in $G$ has at most two outgoing edges,
we can see that $G'$ has at most three times the number of nodes as
$G$.  Graph $G'$ may not be reduced, but it provides an upper bound on the size of a CBDD relative to that of a CZDD\@.

\begin{figure}
\begin{center}
\begin{picture}(5.0,4.75)
\put(0.00,4.50){(A) ZDD/CZDD representation}
\put(0.40,0.00){\includegraphics[scale=0.50]{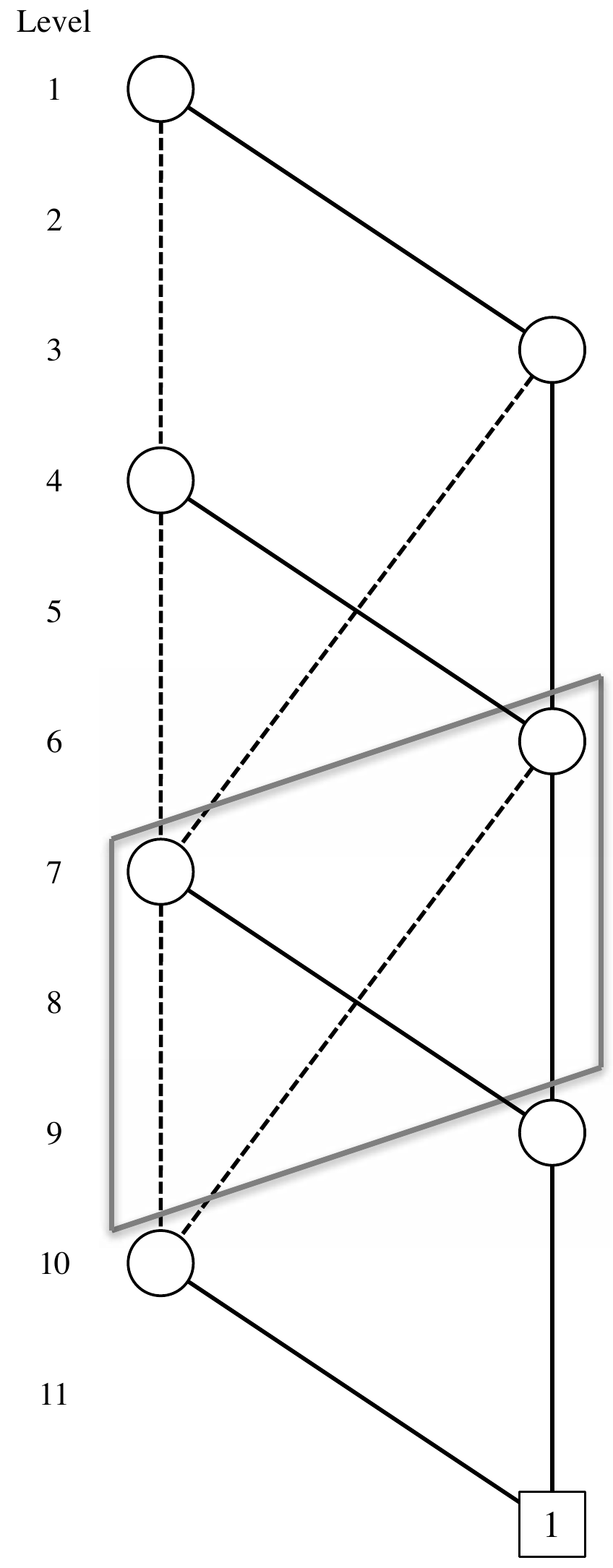}}
\put(2.50,4.50){(B) CBDD representation}
\put(2.90,0.00){\includegraphics[scale=0.50]{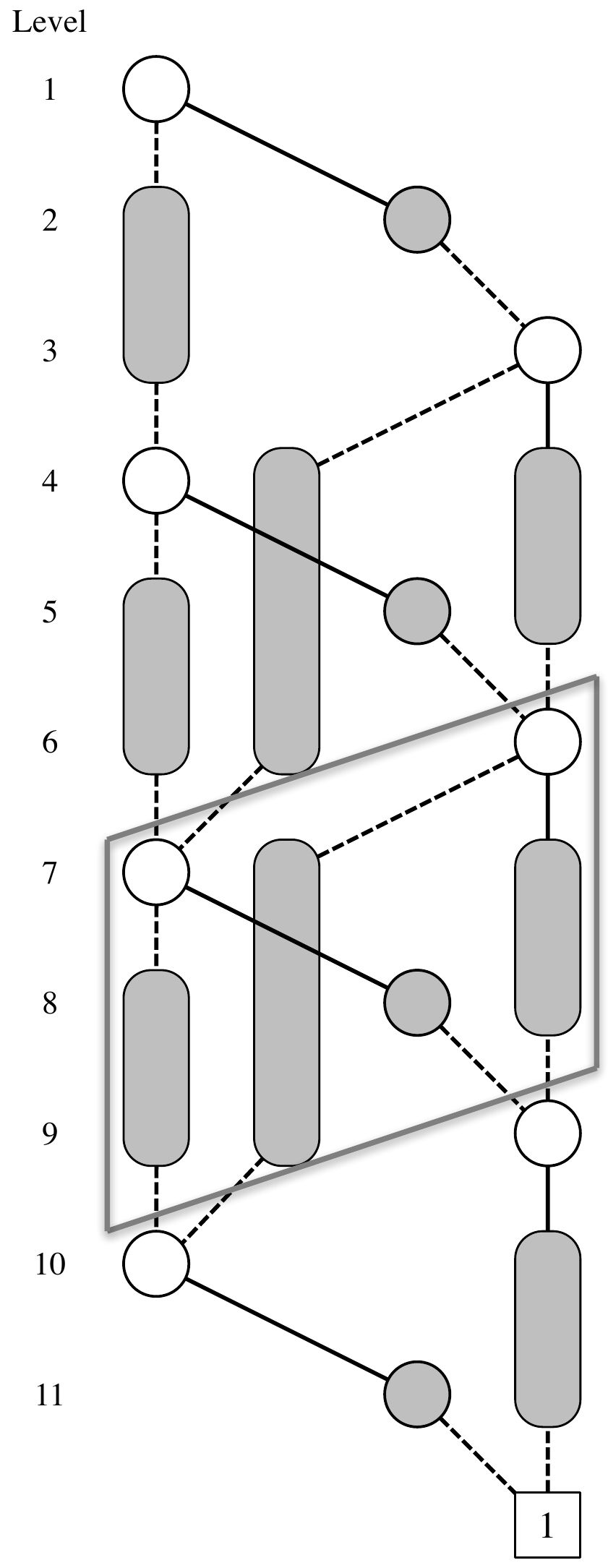}}
\end{picture}
\end{center}
\caption{{\bf Worst case example for effectiveness of CBDD compression}.
The implicit \ztype{} chains in the ZDD (A) must be explicitly encoded in the CBDD (B), increasing its size by a factor of 3.}
\label{fig:cbdd-zdd}
\end{figure}

This bound it tight---Figure \ref{fig:cbdd-zdd} illustrates the
representations for a family of functions, parameterized by a value
$k$ ($k=3$ in the example), such that the function is defined over
$3k+2$ variables.  The ZDD and CZDD representations are identical (A),
having $2k+3$ nodes (including both leaf nodes.)  The CBDD
representation has $6k+2$ nodes (B).  We can see in this example that
the CBDD requires nodes (shown in gray) to encode the \ztype{} chains that
are implicit in the ZDD\@.  To construct the graphs for different
values of $k$, the graph pattern enclosed in the diagonal boxes can
be replicated as many times as are needed.

Second, let us consider Equation \ref{eqn:czdd-bdd}, bounding the
relative sizes of the CZDD and BDD representations of a function.
Consider a graph $G$ representing function $f$ as a BDD\@.  We can
construct its representation $G'$ as a CZDD\@.  Consider each edge $G$
from node $u$, having level $l_u$ to node $v$, having level $l_v$.
Let $r = \lo(v)$ and $s = \hi(v)$.  $G'$ has a node $w_{uv}$ of
the form $\bnode{\lpair{l_u+1}{l_v}}{w_{vr}}{w_{vs}}$.  That is,
$w_{uv}$ encodes any \dctype{} chain between $u$ and $v$, and it
has edges to the nodes generated to encode the edges between $v$
and its two children.  The size of $G'$ is bounded by the number
of edges in $G$, which is at most twice the number of nodes.

\begin{figure}
\begin{center}
\begin{picture}(5.0,3.30)
\put(0.00,3.05){(A) BDD representation}
\put(0.40,0.00){\includegraphics[scale=0.50]{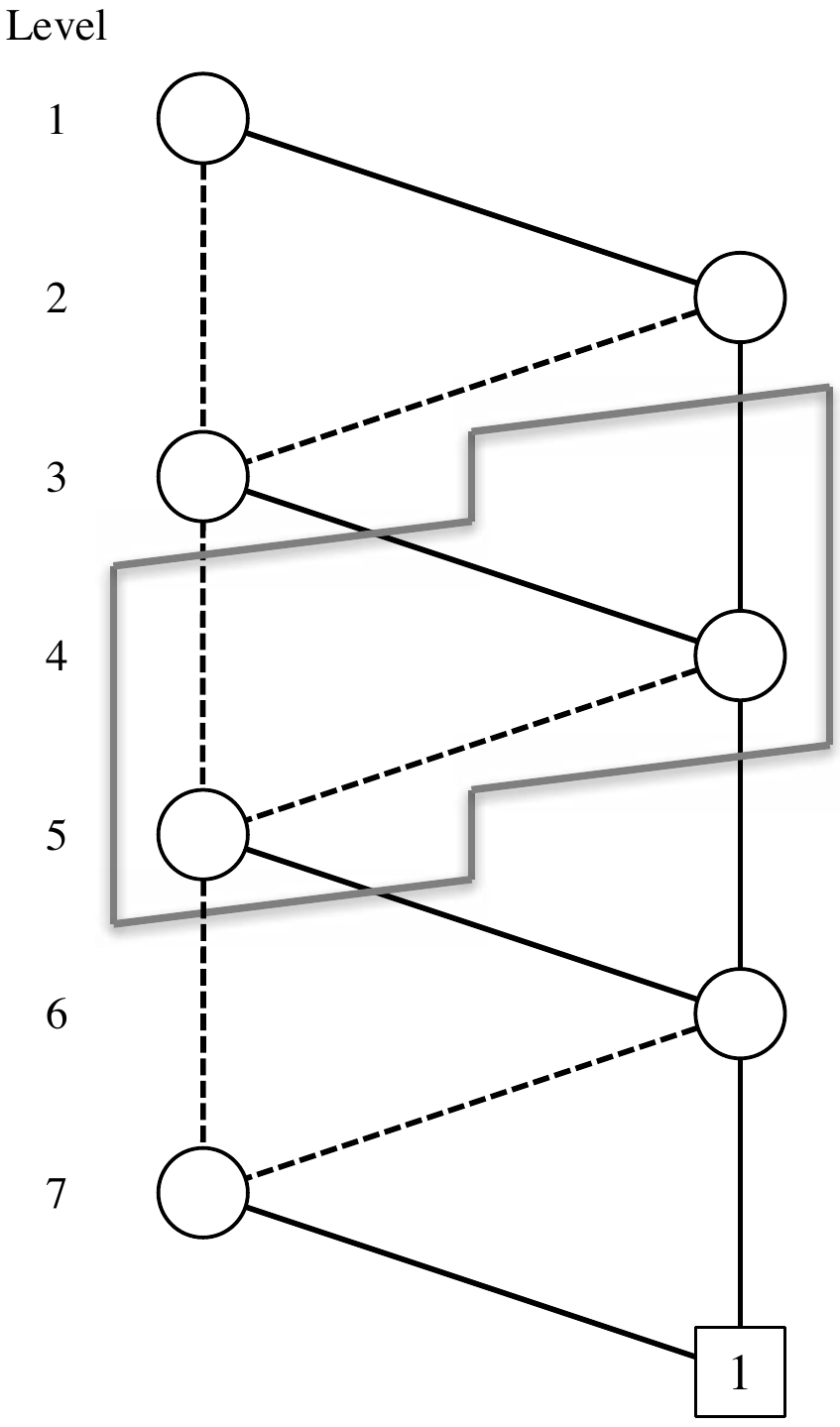}}
\put(2.50,3.05){(B) CZDD representation}
\put(2.90,0.00){\includegraphics[scale=0.50]{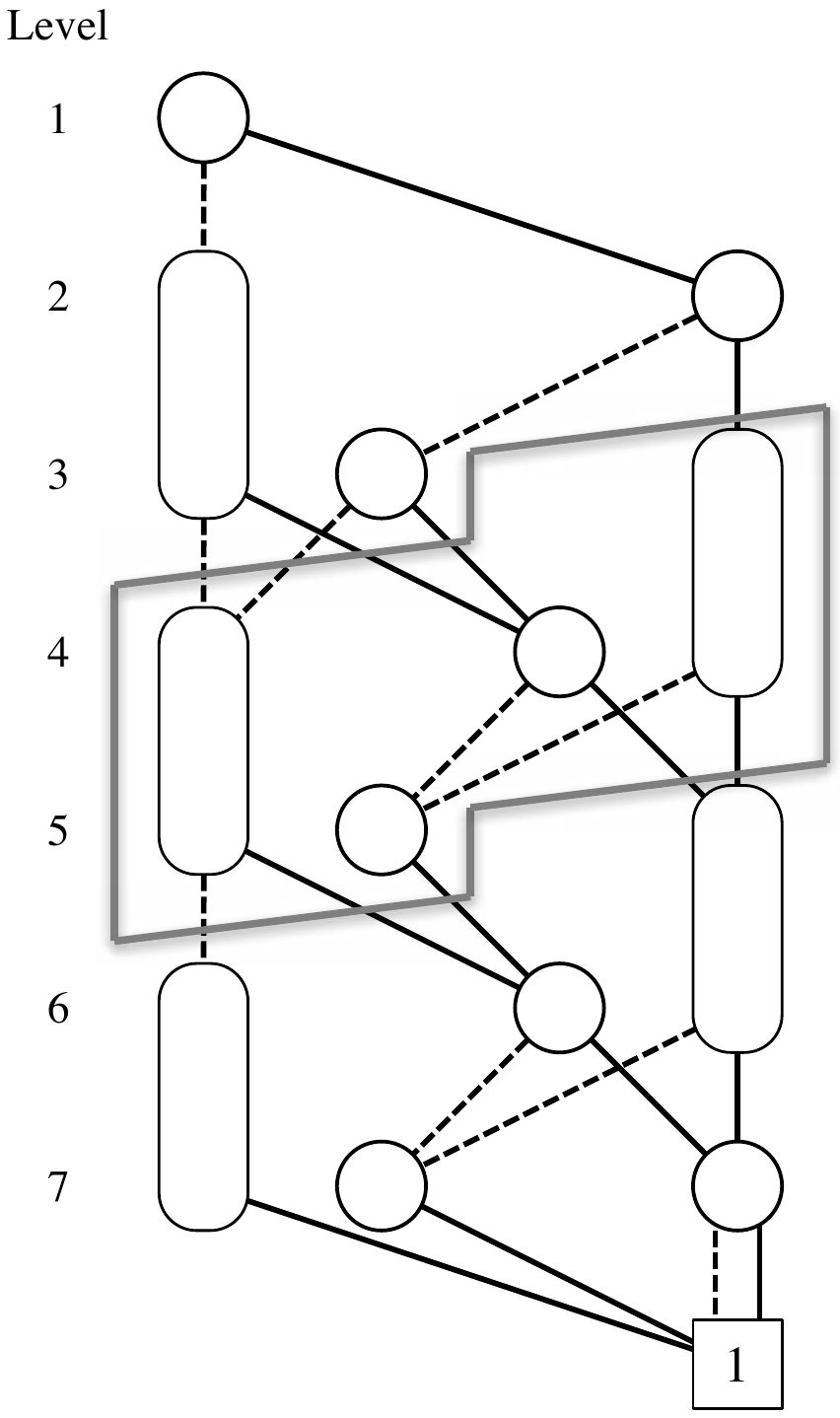}}
\end{picture}
\end{center}
\caption{{\bf Worst case example for effectiveness of CZDD compression}.
The nodes in the BDD (A) must be duplicated to encode the incoming \dctype{} chains (B), increasing the size by a factor of 2.}
\label{fig:czdd-bdd}
\end{figure}

This bound it also tight---Figure \ref{fig:czdd-bdd} illustrates the
representations for a family of functions, parameterized by a value
$k$ ($k=3$ in the example), such that the function is defined over
$2k+1$ variables.  The BDD representations (A) has
$2k+3$ nodes (including both leaf nodes.)  The CZDD
representation has $4k+3$ nodes (B).  We can see that most of the nodes in the BDD must be duplicated: once with no incoming \dctype{} chain, and once with a chain of length one.
To construct the graphs for different
values of $k$, the graph pattern enclosed in the diagonal boxes can
be replicated as many times as are needed.

As can be seen in Figure \ref{fig:relations}, these bounds contain an
asymmetry between BDDs and ZDDs and their compressed forms.  The bound
of 3 holds between CBDDs and CZDDs, and hence by transitivity between
CBDDs and ZDDs, while the bound of 2 holds only between CZDDs and
BDDs.  The general form of the \ortype{} chain (Figure
\ref{fig:chain-patterns}(A)), where $g$ is something other than
$\zero$, cannot be directly encoded with CZDD nodes.

\section{Operating on CBDDs and CZDDs}

The \applyop{} algorithms for decision diagrams operate by
recursively expanding a set of argument decision diagrams according to
a Shannon expansion of the represented functions \cite{bryant:tc86,bryant:hmc17}.  These algorithms
allow functions to be combined according to standard binary Boolean
operations, such as $\andop$, $\orop$, and $\xorop$, as well as by the
if-then-else operation $\iteop$\@.

As notation, consider a step
that expands
$k$ argument nodes $\{v_i | 1 \leq i \leq k\}$ where $v_i = \bnode{\lpair{t_i}{b_i}}{f_i}{g_i}$.
For example, operations $\andop$, $\orop$, and $\xorop$ use the \applyop{} algorithm with
$k=2$, while ternary operations, such as $\iteop$ use $k=3$.
A step can be summarized as follows:
\begin{enumerate}
\item If one of the terminal cases apply, then return the result.
\item If the computed cache contains an entry for this combination of
  operation and arguments, then return the previously computed result.
\item Recursively compute the result:
\begin{enumerate}
\item Choose splitting level(s) based on the levels of the
  arguments.
\item Generate $\hi$ and $\lo$ cofactors for each argument.
\item Recursively compute the $\hi$ and $\lo$ values of the result using the
  \applyop{} algorithm with the $\hi$ cofactors and the $\lo$ cofactors, respectively.
\item Determine the result node parameters based on the computed $\hi$
  and $\lo$ cofactors and the splitting level(s).
\item Either reuse an existing node or create a new one with the
  desired level(s) and $\hi$ and $\lo$ children.
\end{enumerate}
\item Store an entry in the computed cache.
\item Return the computed value.
\end{enumerate}
In generalizing from conventional BDDs and ZDDs to their chained
versions, we need only modify 3(a) (splitting), 3(b) (cofactoring),
and 3(d) (combining) in this sequence, considering the two levels for each node.

For CBDDs, we define the splitting levels $t$ and $b$ as:
\begin{eqnarray}
t & = & \min_{1 \leq i \leq k} t_i \label{eqn:bdd-split} \\
b & = & \min_{1 \leq i \leq k} \left \{
\begin{array}{ll}
b_i, & t_i = t \\
t_i, & t_i = n+1 \\
t_i - 1, & {\rm else} \end{array}
\right . \nonumber
\end{eqnarray}

We then define the two cofactors for each
argument node $v_i$, denoted
$\lo(v_i, \lpair{b}{t})$ and $\hi(v_i, \lpair{b}{t})$, according to the
following table:
\begin{center}
\begin{tabular}{|c|c|c|c|}
\hline
\makebox[0.75in]{Case} & \makebox[0.75in]{Condition} 
   & \makebox[1.0in]{$\lo(v_i, \lpair{b}{t})$} & \makebox[1.0in]{$\hi(v_i, \lpair{b}{t})$} \\
\hline
1 & $b < t_i$ & $v_i$ & $v_i$ \\
2 & $b = b_i$ & $f_i$ & $g_i$ \\
3 & $t_i \leq b < b_i$ & $\bnode{\lpair{b+1}{b_i}}{f_i}{g_i}$ & $g_i$\\
\hline
\end{tabular}
\end{center}

These three cases can be explained as follows:
\begin{description}
\item[Case 1:] Splitting spans levels less than the top level of
  $v_i$.  Since level-skipping edges encode \dctype{} chains, both cofactors equal the original node.
\item[Case 2:] Splitting spans the same levels as node $v_i$.  The
  cofactors are therefore the nodes given by the outgoing edges.
\item[Case 3:] Splitting spans a subset of the levels covered by
  node $v_i$.  We construct a new node spanning the
  remaining part of the encoded \ortype{} chain for the $\lo$ cofactor and
  have $g_i$ as the $\hi$ cofactor.
\end{description}

Recursive application of the \applyop{} operation on the cofactors
generates a pair of nodes $u_{0}$ and $u_{1}$.
Using the variable levels $t$ and
$b$ defined in Equation \ref{eqn:bdd-split}, these are
combined to form a result node $u$, defined as follows:

\begin{eqnarray}
u & = &  \left  \{
\begin{array}{lll}
u_{0}, & u_{0} = u_{1} & {\rm Case~1}\\ 
\bnode{\lpair{t}{b'}}{w_{0}}{u_{1}}, & u_{0} =
\bnode{\lpair{b+1,b'}}{w_{0}}{u_{1}} & {\rm Case~2}\\
\bnode{\lpair{t}{b}}{u_{0}}{u_{1}}, & {\rm else} & {\rm Case~3} \label{eqn:bdd-form} \\
\end{array} 
\right . 
\end{eqnarray}

These three cases can be explained as follows:
\begin{description}
\item[Case 1:] The $\hi$ and $\lo$ cofactors are identical, and so
  the don't-care reduction rule can be applied.
\item[Case 2:] Chain compression can be applied to create a node that
  absorbs the $\lo$ cofactor.
\item[Case 3:] No special rules apply.
\end{description}

Similar rules hold for applying operations to CZDDs, although there
are important differences, due to the different interpretations of
level-skipping edges.

We define the splitting levels $t$ and $b$ as:
\begin{eqnarray}
t & = & \min_{1 \leq i \leq k} t_i \label{eqn:zdd-split} \\
b & = & \min_{1 \leq i \leq k} \left \{
\begin{array}{ll}
b_i, & t_i = t \\
n+1, & v_i = \zero\\
t, & {\rm else}
\end{array}
\right . \nonumber
\end{eqnarray}

The cofactors for
argument node $v_i$
are defined according to the following table:
\begin{center}
\begin{tabular}{|c|c|c|c|}
\hline
\makebox[0.75in]{Case} &\makebox[0.75in]{Condition} & \makebox[1.0in]{$\lo(v_i, \lpair{b}{t})$} & \makebox[1.0in]{$\hi(v_i, \lpair{b}{t})$} \\
\hline
1 & $b < t_i$ & $v_i$ & $\zero$ \\
2 & $b = b_i$ & $f_i$ & $g_i$ \\
3 & $t_i \leq b < b_i$ & $\znode{\lpair{b+1}{b_i}}{f_i}{g_i}$ & $\znode{\lpair{b+1}{b_i}}{f_i}{g_i}$ \\
\hline
\end{tabular}
\end{center}

These three cases can be explained as follows:
\begin{description}
\item[Case 1:] The splitting spans levels less than the top level of
  $v_i$.  Since level-skipping edges encode \ztype{} chains, the $\lo$
  cofactor equals the original node and the $\hi$ cofactor equals
  leaf $\zero$.
\item[Case 2:] The splitting spans the same levels as node $v_i$.  The
  cofactors are therefore the nodes given by the outgoing edges.
\item[Case 3:] The splitting spans a subset of the levels covered by
  node $v_i$.  We construct a new node spanning the
  remaining part of the encoded \dctype{} chain for both cofactors.
\end{description}

Recursive application of the \applyop{} operation on the cofactors
generates a pair of nodes $u_{0}$ and $u_{1}$.
Using the variable ranges $t$ and
$b$ defined in Equation \ref{eqn:zdd-split}, these are
combined to form a result node $u$, defined as follows:

\begin{eqnarray}
u & = &  \left  \{
\begin{array}{lll}
u_{0}, & u_{1} = \zero \; {\rm and} \; t = b  & {\rm Case~1}\\ 
\bnode{\lpair{t}{b-1}}{u_{0}}{u_{0}}, & u_{1} = \zero \; {\rm and} \; t < b  & {\rm Case~2}\\ 
\bnode{\lpair{t}{b'}}{w_{0}}{w_{1}}, & u_{0} = u_{1} =
\bnode{\lpair{b+1,b'}}{w_{0}}{w_{1}}  & {\rm Case~3}\\ 
\bnode{\lpair{t}{b}}{u_{0}}{u_{1}}, & {\rm else}  & {\rm Case~4} \label{eqn:zdd-form} \\
\end{array} 
\right . 
\end{eqnarray}

These four cases can be explained as follows:
\begin{description}
\item[Case 1:] The zero-suppression rule can be applied to return a
  direct pointer to $u_0$
\item[Case 2:] The zero-suppression rule can be applied, but we must
  construct a node encoding the \dctype{} chain between levels $t$
  and $b-1$.
\item[Case 3:] Chain compression can be applied to create a node that
  absorbs the $\lo$ cofactor.
\item[Case 4:] No special rule applies.
\end{description}

\section{Implementation}

We implemented both CBDDs and CZDDs by modifying version 3.0.0 of the CUDD
BDD package \cite{somenzi:sttt01}.  When compiled for 64-bit
execution, CUDD stores a 32-bit field {\tt index} in each node, where
this index is translated into a level according to the variable
ordering.  For our implementation, we split this field into two 16-bit
fields {\tt index} and {\tt bindex} to (indirectly) encode the top and
bottom levels of the node.  Thus, there was no storage penalty for the
generalization to a chained form.

Incorporating chaining required modifications to many parts of the
code, including how node keys are generated for the unique table and
computed cache, how reductions are performed, and how Boolean
operations are applied to functions.  In addition, many of the
commonly used functions, such as computing the number of satisfying
solutions to a function, computing its support, and enumerating a
minterm representation of a function required modifications.

CUDD uses complement edges when representing BDDs
\cite{brace:dac90,minato:dac90}.  Complement edges can potentially
reduce the size of a BDD by a factor of two, invalidating the size
ratio bounds derived in Equations \ref{eqn:bdd-zdd}--\ref{eqn:zdd-bdd}
and \ref{eqn:cbdd-czdd}--\ref{eqn:czdd-bdd}.  For the experimental
results presented in the following section, we therefore use a
representation based on CUDD's support for {\em Algebraic Decision
  Diagrams} (ADDs) \cite{bahar-iccad93}.  ADDs generalize BDDs by
allowing arbitrary leaf values.  Restricting the leaf values to 0 and
1 yields conventional BDDs without complement edges.  We revisit the
use of complement edges in Section \ref{sect:complement}.

\section{Experimental Results}

We chose three different categories of benchmarks to compare the performance
of BDDs, ZDDs, and their chained versions.  One set of benchmarks
evaluated the ability of DDs to represent information in compact form,
a second to evaluate their ability to solve combinatorial
problems, and a third to represent typical digital logic functions.
All experiments were performed on a 4.2~GHz Intel Core i7 processor with 32~GB of memory
running the OS~X operating system.

\subsection{Encoding a Dictionary}

\begin{figure}
\begin{center}
\begin{tabular}{|l|r|rr|rr|}
\hline
\multicolumn{1}{|c|}{}	& \multicolumn{1}{|c|}{Words} &
\multicolumn{1}{|c}{Radix} & \multicolumn{1}{c|}{Length} &
\multicolumn{1}{|c}{One-hot variables}	& \multicolumn{1}{c|}{Binary variables} \\
\hline
Compact word list	&  235,886 	&  54 	&  24 	&  1,296 	&  144  \\
ASCII word list	&  235,886 	&  129 	&  24 	&  3,096 	&  192  \\
Compact password list	&  979,247 	&  80 	&  32 	&  2,560 	&  192  \\
ASCII password list	&  979,247 	&  129 	&  32 	&  4,128 	&  256  \\
\hline
\end{tabular}
\end{center}
\caption{Characteristics of Dictionary Benchmarks}
\label{fig:dict:properties}
\end{figure}

As others have observed \cite{knuth:v4a}, a list of words can be
encoded as a function mapping strings in some alphabet to either 1
(included in list) or 0 (not included in list.)  Strings can further
be encoded as binary sequences by encoding each symbol as a sequence
of bits, allowing the list to be represented as a Boolean function.
We consider two possible encodings of the symbols, defining the {\em
  radix} $r$ to be the number of possible symbols.  A {\em one-hot}
encoding (also referred to as a ``1-of-N'' encoding) requires $r$ bits
per symbol.  Each symbol is assigned a unique position, and the symbol
is represented with a one in this position and zeros in the rest.  A
{\em binary} encoding requires $\lceil \log_2 r \rceil$ bits per
symbol.  Each symbol is assigned a unique binary pattern, and the
symbol is represented by this pattern.  Lists consisting of words with
multiple lengths can be encoded by introducing a special ``null''
symbol to terminate each word.

Figure \ref{fig:dict:properties} shows the characteristics of eight
benchmarks derived from two word lists to allow comparisons of
different encoding techniques and representations.  The first list is the
file {\tt /usr/share/dict/words} found on Macintosh systems.  It is
derived from the word list used by the original Unix spell checker.
It contains 235,886 words with lengths ranging from one to 24 symbols,
and where the symbols consist of lower- and upper-case letters plus
hyphen.  We consider two resulting symbol sets: a {\em compact} form,
consisting of just the symbols found in the words plus a null symbol
(54 total), and an {\em ASCII} form, consisting of all 128 ASCII
characters plus a null symbol.

The second word list is from an online list of words employed by
password crackers.  It consists of 979,247 words ranging in length
from one to 32 symbols, and where the symbols include 79 possible
characters.  Again, we consider both a compact form and an ASCII form.

As Figure \ref{fig:dict:properties} shows, the choice of one-hot
vs.~binary encoding has a major effect on the number of Boolean
variables required to encode the words.  With a one-hot encoding, the
number of variables ranges between 1,296 and 4,128, while 
it ranges between 144 and 256 with a binary representation.

To generate DD encodings of a word list, we first constructed a trie
representation the words and then generated Boolean formulas via a
depth-first traversal of the trie.

\begin{figure}
\begin{center}
\begin{tabular}{|l|rrr|rr|}
\cline{2-6}
\multicolumn{1}{l|}{{\bf One-hot}} & 
\multicolumn{3}{|c|}{Node counts} &
\multicolumn{2}{|c|}{Ratios} \\
\cline{2-6}
\multicolumn{1}{c|}{} & 
\multicolumn{1}{|c}{BDD} & \multicolumn{1}{c}{CBDD}	&
\multicolumn{1}{c|}{(C)ZDD} & 
\multicolumn{1}{|c}{BDD:CBDD}	& \multicolumn{1}{c|}{CBDD:CZDD} \\
\hline
Compact word list	&  9,701,439 	&  626,070 	&  297,681 	& 15.50	& 2.10	\\
ASCII word list	&  23,161,501 	&  626,071 	&  297,681 	& 37.00	& 2.10	\\
Compact password list	&  49,231,085 	&  2,321,572 	&  1,130,729  	& 21.21	& 2.05	\\
ASCII password list	&  79,014,931 	&  2,321,792 	&  1,130,729  	& 34.03	& 2.05	\\
\hline
\multicolumn{6}{c}{} \\
\cline{2-6}
\multicolumn{1}{l|}{{\bf Binary}} &
\multicolumn{3}{|c|}{Node counts} &
\multicolumn{2}{|c|}{Ratios} \\
\cline{2-6}
\multicolumn{1}{c|}{} & 
\multicolumn{1}{|c}{BDD} & \multicolumn{1}{c}{CBDD}	&
\multicolumn{1}{c|}{(C)ZDD} &
\multicolumn{1}{|c}{BDD:CBDD}	& \multicolumn{1}{c|}{CBDD:CZDD} \\
\hline
Compact word list	&  1,117,454 	&  1,007,868 	&  723,542 	& 1.11	& 1.39\\
ASCII word list	&  1,464,773 	&  1,277,640 	&  851,580 	& 1.15	& 1.50\\
Compact password list	&  4,422,292 	&  3,597,474 	&  2,506,088 	&   1.23	& 1.44	\\
ASCII password list	&  4,943,940 	&  4,307,614 	&  2,875,612 	&   1.15	& 1.50	\\
\hline
\end{tabular}
\end{center}
\caption{Node counts and ratios of node counts for dictionary benchmarks}
\label{fig:dict:counts}
\end{figure}

Figure \ref{fig:dict:counts} shows the number of nodes required to
represent word lists as Boolean functions, according to the different
lists, encodings, and DD types.  The entry labeled ``(C)ZDD'' gives the node counts for both ZDDs and CZDDs.
These are identical, because
there were no \dctype{} chains for these functions.
The two columns on the right show the ratios
between the different DD types.  Concentrating first on one-hot
encodings, we see that the chain compression of CBDDs reduces the size
compared to BDDs by large factors (15.50--34.03).  Compared to ZDDs,
representing the lists by CBDDs incurs some penalty (2.05--2.10), but
less than the worst-case penalty of 3.  
Increasing the radix from a compact form to the full
ASCII character set causes a significant increase in BDD size, but
this effect is eliminated by using the zero suppression capabilities
of CBDDs, ZDDs, and CZDDs.

Using a binary encoding of the symbols reduces the variances between
the different encodings and DD types.  CBDDs provide only a small
benefit (1.11--1.23) over BDDs, and CBDDs are within a factor of 1.50
of ZDDs.  Again, chaining of ZDDs provides no benefit.  Observe that
for both lists, the most efficient representation is to use either
ZDDs or CZDDs with a one-hot encoding.  The next best is to use CBDDs
with a one-hot encoding, and all three of these are insensitive to
changes in radix.  These cases demonstrate the ability of ZDDS (and
CZDDs) to use very large, sparse encodings of values.  By using
chaining, CBDDs can also take advantage of this property.

\begin{figure}
\begin{center}
\begin{tabular}{|l|rrr|rrr|}
\cline{2-7}
\multicolumn{1}{l|}{{\bf One-hot}} & 
\multicolumn{3}{|c|}{Operations} &
\multicolumn{3}{|c|}{Time (secs.)} \\
\cline{2-7}
\multicolumn{1}{c}{} & 
\multicolumn{1}{|c}{ZDD} & \multicolumn{1}{c}{CZDD}	&
\multicolumn{1}{c|}{Ratio} & 
\multicolumn{1}{|c}{ZDD} & \multicolumn{1}{c}{CZDD}	&
\multicolumn{1}{c|}{Ratio} \\
\hline
Compact word list &  142,227,877  &  12,097,435  & 1.38 & 48.78 & 15.04 & 3.24 \\
ASCII word list &  375,195,184  &  28,574,814  & 13.13 & 173.56 & 21.84 & 7.95 \\
Compact password list &  806,017,001  &  62,785,274  & 12.84 & 713.15 & 46.73 & 15.26 \\
ASCII password list &  1,383,534,557  &  104,059,626  & 13.30 & 658.21 & 57.81 & 11.39 \\
\hline
\multicolumn{7}{c}{} \\
\cline{2-7}
\multicolumn{1}{l|}{{\bf Binary}} & 
\multicolumn{3}{|c|}{Operations} &
\multicolumn{3}{|c|}{Time (secs.)} \\
\cline{2-7}
\multicolumn{1}{c}{} & 
\multicolumn{1}{|c}{ZDD} & \multicolumn{1}{c}{CZDD}	&
\multicolumn{1}{c|}{Ratio} & 
\multicolumn{1}{|c}{ZDD} & \multicolumn{1}{c}{CZDD}	&
\multicolumn{1}{c|}{Ratio} \\
\hline
Compact word list &  15,701,738  &  1,826,171  & 8.60 &  13.11 & 9.70 & 1.35 \\
ASCII word list &  20,921,746  &  2,139,574  & 9.78 & 14.40 & 10.20 & 1.41 \\
Compact password list &  66,489,058  &  7,499,615  & 8.87 & 52.52 & 30.62 & 1.72 \\
ASCII password list &  75,556,080  &  7,936,321  & 9.52 &  50.77 & 30.33 & 1.67 \\
\hline
\end{tabular}
\end{center}
\caption{Impact of chaining on effort required to generate DD representations of word lists.}
\label{fig:dict:effort}
\end{figure}

Although the final node counts for the benchmarks indicate no
advantage of chaining for ZDDs, statistics characterizing the effort
required to derive the functions show a significant benefit.
Figure \ref{fig:dict:effort} indicates the total number of
operations and the total time required for generating ZDD and CZDD
representations of the benchmarks.  The operations are computed as the
number of times the program checks for an entry in the operation
cache (step 2 in the description of the \applyop{} algorithm). 
There are many operational factors that can
affect the number of operations, including the program's policies for
operation caching and garbage collection.  Nevertheless, it is some
indication of the amount of activity required to generate the DDs.  We
can see that chaining reduces the number of operations by factors of
8.87--13.30.  The time required depends on many attributes of the DD
package and the system hardware and software.  Here we see that
chaining improves the execution time by factors of 1.35--15.26.

With unchained ZDDs, many of the intermediate functions have large
\dctype{} chains.  For example, the ZDD representation of the function
$x$, for variable $x$, requires $n+2$ nodes---one for the variable,
$n-1$ for the \dctype{} chains before and after the variable, and two
leaf nodes.  With chaining, this function reduces to just four nodes:
the upper \dctype{} chain is incorporated into the node for the
variable, and a second node encodes the lower chain.  Our dictionary
benchmarks have as many as 4,000 variables, and so some of the
intermediate DDs can be 1,000 times more compact due to
chaining.

\subsection{The 15-Queens Problem}

A second set of benchmarks involved encoding all possible solutions to
the $n$-queens problem.  This problem attempts to place $n$ queens on
a $n\times n$ chessboard in such a way that no two queens can attack
each other.  For our benchmark we chose $n = 15$ to stay within the
memory limit of the processor being used.  Once a DD representation
has been generated for all possible solutions, it is easy to then
compute useful properties, such as the number of solutions.

Once again, there are two choices for encoding the positions of queens
on the board.  A {\em one-hot} encoding uses a Boolean variable for
each square.  A {\em binary} encoding uses $\lceil \log_2 n \rceil = 4$ variables for
each row, encoding the position of the queen within the row.

The Boolean formulas  must then impose the following constraints:
\begin{description}
\item[{\bf ROW}:] No two squares in the same row are occupied.  There
  are $n$ such constraints.  A binary encoding already imposes this constraint.
\item[{\bf COL}:] No two squares in the same column are occupied.
  There are $n$ such constraints.
\item[{\bf DIAG}:] No two squares in the same diagonal from upper left
  to lower right are occupied.  There are $2n-1$ such constraints.
\item[{\bf OFF-DIAG}:] No two squares in the same diagonal from upper
  right to lower left are occupied.  There are $2n-1$ such constraints.
\end{description}
Our most successful approach worked from the bottom row to the top.
At each level, it generated formulas for each column, diagonal, and
off-diagonal expressing whether it was occupied in the rows at or
below this one, based on the formulas for the level below and the
variables for the present row.  

We considered two ways of ordering the variables for the different
rows.  The {\em top-down} ordering listed the variables according to
the row numbers 1 through 15.  The {\em center-first} ordering listed
variables according to the row number sequence
$8,9,7,10,6,11,5,12,4,13,3,14,2,15,1$.  Our hope in using this sequence
was that ordering the center rows first would reduce the DD
representation size.  This proved not to be the case, but the
resulting node counts are instructive.

\begin{figure}
\begin{center}
\begin{tabular}{|ll|rrr|rr|}
\cline{3-7}
\multicolumn{2}{l|}{{\bf One-hot}} & 
\multicolumn{3}{|c|}{Node counts} &
\multicolumn{2}{|c|}{Ratios} \\
\hline
\multicolumn{1}{|c}{Ordering} & \multicolumn{1}{c|}{Graph(s)} & 
\multicolumn{1}{|c}{BDD} & \multicolumn{1}{c}{CBDD}	&
\multicolumn{1}{c|}{CZDD} & 
\multicolumn{1}{|c}{BDD:CBDD}	& \multicolumn{1}{c|}{CBDD:CZDD} \\
\hline
Top-down & Final &  51,889,029  &  10,529,738  &  4,796,504  & 4.93 & 2.20  \\
Top-down & Peak &  165,977,497  &  39,591,936  &  18,625,659  & 4.19 & 2.13  \\
Center-first & Final &  65,104,658  &  12,628,086  &  5,749,613  & 5.16 & 2.20  \\
Center-first & Peak &  175,907,712  &  42,045,602  &  19,434,105  & 4.18 & 2.16  \\
\hline
\multicolumn{6}{c}{} \\
\cline{3-7}
\multicolumn{2}{l|}{{\bf Binary}} & 
\multicolumn{3}{|c|}{Node counts} &
\multicolumn{2}{|c|}{Ratios} \\
\hline
\multicolumn{1}{|c}{Ordering} & \multicolumn{1}{c|}{Graph(s)} & 
\multicolumn{1}{|c}{BDD} & \multicolumn{1}{c}{CBDD}	&
\multicolumn{1}{c|}{CZDD} &
\multicolumn{1}{|c}{BDD:CBDD}	& \multicolumn{1}{c|}{CBDD:CZDD} \\
\hline
Top-down & Final &  13,683,076  &  11,431,403  &  7,383,739  & 1.20 & 1.55  \\
Top-down & Peak &  43,954,472  &  38,898,146  &  26,682,980  & 1.13 & 1.46  \\
Center-first & Final &  17,121,947  &  14,185,276  &  9,054,115  & 1.21 & 1.57  \\
Center-first & Peak &  46,618,943  &  41,362,659  &  28,195,596  & 1.13 & 1.47  \\
\hline
\end{tabular}
\end{center}
\caption{Node counts and ratios of node counts for 15-queens benchmarks}
\label{fig:queens:counts}
\end{figure}

Figure \ref{fig:queens:counts} shows the node counts for the different
benchmarks.  It shows both the size of the final function representing
all solutions to the 15-queens problem, as well as the peak size,
computed as the maximum across all rows of the combined size of the
functions that are maintained to express the constraints imposed by
the row and those below it.  For both the top-down and the center-first benchmarks, this maximum
was reached after completing row 3.
Typically the peak
size was around three times larger than the final size.

For a one-hot encoding, we can see that CBDDs achieve factors of
4.18--5.16 compaction over BDDs, and they come within a factor of
2.20 of CZDDs.  For a binary encoding, the levels of compaction are
much less compelling (1.13--1.20), as is the advantage of CZDDs over
BDDs.  It is worth noting that the combination of a one-hot encoding
and chaining gives lower peak and final sizes than BDDs with a binary encoding.

\begin{figure}
\begin{center}
\begin{tabular}{|ll|rr|r|}
\cline{3-5}
\multicolumn{2}{l|}{{\bf One-hot}} & 
\multicolumn{2}{|c|}{Node counts} &
\multicolumn{1}{|c|}{Ratios} \\
\hline
\multicolumn{1}{|c}{Ordering} & \multicolumn{1}{c|}{Graph(s)} & 
\multicolumn{1}{|c}{ZDD} & \multicolumn{1}{c|}{CZDD}	&
\multicolumn{1}{|c|}{ZDD:CZDD} \\
\hline
Top-down & Final &  4,796,504  &  4,796,504  & 1.00  \\
Top-down & Peak &  18,632,019  &  18,625,659  & 1.00  \\
Center-first & Final &  5,749,613  &  5,749,613  & 1.00  \\
Center-first & Peak &  73,975,637  &  19,434,105  & 3.81  \\
\hline
\multicolumn{5}{c}{} \\
\cline{3-5}
\multicolumn{2}{l|}{{\bf Binary}} & 
\multicolumn{2}{|c|}{Node counts} &
\multicolumn{1}{|c|}{Ratios} \\
\hline
\multicolumn{1}{|c}{Ordering} & \multicolumn{1}{c|}{Graph(s)} & 
\multicolumn{1}{|c}{ZDD} & \multicolumn{1}{c|}{CZDD}	&
\multicolumn{1}{|c|}{ZDD:CZDD} \\
\hline
Top-down & Final &  7,383,739  &  7,383,739  & 1.00  \\
Top-down & Peak &  26,684,315  &  26,682,980  & 1.00  \\
Center-first & Final &  9,054,115  &  9,054,115  & 1.00  \\
Center-first & Peak &  33,739,362  &  28,195,596  & 1.20 \\
\hline
\end{tabular}
\end{center}
\caption{Effect of chaining for ZDD representations of 15-queens benchmarks}
\label{fig:queens:zdd}
\end{figure}

Figure \ref{fig:queens:zdd} compares the sizes of the ZDD and CZDD
representations of the 15-queens functions.  We can see that the final
sizes are identical---there are no \dctype{} chains in the functions
encoding problem solutions.  For the top-down ordering, CZDDs also
offer only a small advantage for the peak requirement.  For the
center-first ordering, especially with a one-hot encoding, however, we
can see that CZDDs are significantly (3.81$\times$) more compact.  As
the construction for row 3 completes, the variables that will encode
the constraints for rows 2 and 5 remain unconstrained, yielding many
\dctype{} chains.  Once again, this phenomenon is much smaller with a
binary encoding.  In the end, the center-first variable ordering does
not outperform the more obvious, top-down ordering, but it was
beneficial to be able to test it.  By using chaining, CZDDs can
mitigate the risk of trying a nonoptimal variable order.

By way of comparison, Kunkle, Slavici, and Cooperman
\cite{kunkle:pasco10} also used the $n$-queens problem to test their
BDD implementation performed on a cluster of 64 machines.  They used a
one-hot encoding, but generated the formula by forming a conjunction
of all of the constraints on each single position on the board,
requiring $n^2$ constraints, each with up to $4n-4$ terms, requiring a
total of $\theta(n^3)$ Boolean operations.  In constructing a representation
of the 15-queens problem, their program reached a peak of 917,859,646
nodes, a factor of 5.5 times greater than our peak.  Our row-by-row
constraint generation requires only $\theta(n^2)$ Boolean operations, and it
significantly reduces the peak memory requirement.  Minato
described a third approach for generating constraints for the $n$-queens problem, but his
method also requires $\theta(n^3)$ operations \cite{minato:book95}.

\subsection{Digital Circuits}

\begin{figure}
\begin{center}
\begin{tabular}{|l|rrr|rr|}
\cline{2-6}
\multicolumn{1}{c|}{} &
\multicolumn{3}{|c|}{Node counts} &
\multicolumn{2}{|c|}{Ratios} \\
\hline
\multicolumn{1}{|c|}{Circuit} & 
\multicolumn{1}{|c}{BDD} & \multicolumn{1}{c}{ZDD}	&
\multicolumn{1}{c|}{CZDD} & 
\multicolumn{1}{|c}{ZDD:BDD}	& \multicolumn{1}{c|}{CZDD:BDD} \\
\hline
c432	&	 31,321 	&	 48,224 	&	 41,637 	&	1.54	&	1.33 \\
c499	&	 49,061 	&	 49,983 	&	 48,878 	&	1.02	&	1.00 \\
c880	&	 23,221 	&	 52,436 	&	 32,075 	&	2.26	&	1.38 \\
c1908	&	 17,391 	&	 18,292 	&	 17,017 	&	1.05	&	0.98 \\
c2670	&	 67,832 	&	 261,736 	&	 85,900 	&	3.86	&	1.27 \\
c3540	&	 3,345,341 	&	 4,181,847 	&	 3,538,982 	&	1.25	&	1.06 \\
c5315	&	 636,305 	&	 898,912 	&	 681,440 	&	1.41	&	1.07 \\
c6288	&	 48,181,908 	&	 48,331,495 	&	 48,329,117 	&	1.00	&	1.00 \\
c7552   &        4,537          &        37,689         &        4,774          &       8.31    &       1.05 \\
\hline
\end{tabular}
\end{center}
\caption{Node counts and ratios of node counts for digital circuit benchmarks}
\label{fig:circuits:counts}
\end{figure}

BDDs are commonly used in digital circuit design automation, for such
tasks as verification, test generation, and circuit synthesis.  There
ability to represent functions typically found in digital circuits has
been widely studied.  It is natural to study the suitability of
ZDDs, as well as chained versions of BDDs and ZDDs for these applications.

\hyphenation{ISCAS}

We selected the circuits from the ISCAS~'85 benchmark
suite \cite{brglez:iscas85}.  These were originally developed as
benchmarks for test generation, but they have also been widely used as
benchmarks for BDDs \cite{fujita:iccad88,malik:iccad88}.  We generated
variable orderings for all but last two benchmarks by traversing the
circuit graphs, using the fanin heuristic of \cite{malik:iccad88}.
Circuit c6288 implements a $16 \times 16$ multiplier.  For this
circuit, the ordering of inputs listed in the file provided a feasible
variable ordering, while the one generated by traversing the circuit
exceeded the memory limits of our machine.  For c7552, neither the
ordering in the file, nor that provided by traversing the graph,
generated a feasible order.  Instead, we manually generated an
ordering based on our analysis of a reverse-engineered version of the
circuit described in \cite{hansen:ieeedt99}.

Figure \ref{fig:circuits:counts} presents data on the sizes of the DDs
to represent all of the circuit outputs.  We do not present any data
for CBDDs, since these were all close in size to BDDs.  We
can see that the ZDD representations for these circuits are always
larger than the BDD representations, by factors ranging up to 8.31.
Using CZDDs mitigates that effect, yielding a maximum size ratio of
1.38.

Circuit c6288 has historical interest.  Integer multiplication is
known to be intractable for BDDs regardless of variable ordering
\cite{bryant:tc91}.  As a result, this benchmark was long considered
out of reach for a BDD representation and therefore typically omitted
from many benchmark comparisons.  With a modern machine, the benchmark
is achievable, requiring a peak of around 16~GB to represent the data
structures. On the other hand, we see that these functions contain
very few \dctype{} or \ortype{} chains, and hence all four DD types
require around 48 million nodes to represent them.

\subsection{Observations}

These experiments, while not comprehensive, demonstrate that
chaining can allow BDDs to make use of large, sparse encodings, one of
the main strengths of ZDDs.  They also indicate that CZDDs may be the
best choice overall.  They have all of the advantages of ZDDs, while
avoiding the risk of intermediate functions growing excessively large
due to \dctype{} chains.  They are guaranteed to stay within a factor
of 2$\times$ of BDDs.  Even for digital circuit functions, we found
this bound to be conservative---all of the benchmarks stayed within a
factor of 1.4$\times$.

Experienced ZDD users take steps to avoid \dctype{} chains, for
example, by implementing special algorithms that operate directly on
ZDDs, rather than expressing a computation as a sequence of Boolean
operations \cite{knuth:v4a,yoneda:fmcad96}.  By using chaining, CZDDs reduce the cost of these chains,
enabling users to express their computations at the more abstract
level of Boolean expressions, rather than implementing special
algorithms.

\section{Complement Edges}
\label{sect:complement}

As mentioned earlier, CUDD uses complement edges in its representation
of BDDs, and
hence our experiments showed the results for ADDs.  We also
implemented CBDDs with complement edges by modifying
CUDD's implementation of BDDs.
The standard rules for deciding how to
canonicalize complement edges \cite{brace:dac90} can be used without modification with
CBDDs.
Interestingly, these rules imply that there are no \ztype{}
chains in our CBDDs, because the canonicalization forbids having a
$\hi$ pointer to leaf $\zero$.  Instead, such chains are represented by
chain nodes with the $\hi$ edges pointing to leaf $\one$, and with
their incoming edges complemented.  Although Minato's original paper
on ZDDs \cite{minato:dac90} includes a set of conventions for using
complement edges with ZDDs, these are not implemented in CUDD\@.
Thus, we did not attempt this feature in our implementation.

\begin{figure}
\begin{center}
\begin{tabular}{|l|rr|r|}
\cline{2-3}
\multicolumn{1}{c|}{} & \multicolumn{2}{|c|}{Complement edges} &
\multicolumn{1}{|c}{} \\
\hline
\multicolumn{1}{|c|}{Metric} &
\multicolumn{1}{|c}{No} &
\multicolumn{1}{c|}{Yes} &
\multicolumn{1}{|c|}{Ratio} \\
\hline
Final size	&	 48,177,349 	&	 41,417,996 	&	1.16	\\
Operations	&	 675,645,812 	&	 272,783,843 	&	2.48	\\
Time (secs.)	&	1757	&	280	&	6.27	\\
\hline
\end{tabular}
\end{center}
\caption{Benefits of complement edges for CBDD representations of c6288 benchmark.}
\label{fig:complement}
\end{figure}

Figure \ref{fig:complement} illustrates the impact of complement edges
for CBDDs, based on the c6288 benchmark.  Similar results hold without
chaining.  As can be seen, although complement edges can potentially
yield a 2$\times$ reduction in the number of nodes, the actual
reduction is much smaller (1.16).  However, it can greatly affect the
number of operations required (2.48$\times$), since complement edges
enable complementing a function in a single step, rather than
traversing the graph. (The circuit consists mostly of \opname{nor}
gates, each requiring a complement operation.) The impact on time
(6.27$\times$) is even greater, since the large number of complement
operations pollutes the operation cache.

\section{Further Work}

Our modifications to CUDD to support chaining were only sufficient to
evaluate the basic concepts.  Fully integrating these changes into
such a complex software would require significantly more effort.
Perhaps the most challenging would be to implement dynamic variable
ordering \cite{rudell:iccad93}.  The same basic principles of dynamic
variable ordering hold for both CBDDs and CZDDs.  An exchange of
variables at levels $l$ and $l+1$ could be performed without modifying
any of the nodes with levels less than $l$ and without modifying any
external node pointers.  Special consideration must be given to nodes
with levels $\lpair{t}{b}$, such that either $t = l+1$ or $b = l$.
Many low-level details and heuristics of the
implementation would need to be altered to enable dynamic variable
ordering to work well. 

\section{Acknowledgements}

This work has benefitted from conversations with Shin-Ichi Minato and Ofer Strichman.

\end{document}